\documentclass[fleqn,usenatbib]{mnras}

\usepackage{newtxtext,newtxmath}
\usepackage[T1]{fontenc}

\DeclareRobustCommand{\VAN}[3]{#2}
\let\VANthebibliography\thebibliography
\def\thebibliography{\DeclareRobustCommand{\VAN}[3]{##3}\VANthebibliography}

\usepackage{graphicx}	% Including figure files
\usepackage{amsmath}	% Advanced maths commands
\usepackage{amssymb}	% Extra maths symbols

\usepackage{bm}
\usepackage{color}
\usepackage{ulem}
\usepackage{xspace}
\usepackage{float}
\usepackage{graphicx}
\usepackage{epstopdf}
\usepackage{CJKutf8}

\ttfamily
\definecolor{mygray}{gray}{0.6}
\definecolor{magenta}{rgb}{0.858, 0.188, 0.478}
\definecolor{revise}{RGB}{0, 139, 0 }
\definecolor{revise2}{RGB}{255, 0, 0 }

\newcommand{\xxx}[1]{\textcolor{blue}{\textbf{xxx}\xspace}}
\newcommand{\changed}[1]{#1}
\newcommand{\changedd}[1]{#1}

\newcommand{\fg}[1]{Fig.~\ref{fig:#1}}
\newcommand{\Fg}[1]{Figure~\ref{fig:#1}}%beginning of the sentence

\newcommand{\eq}[1]{Eq.~(\ref{eq:#1})\xspace}
\newcommand{\Eq}[1]{Equation~(\ref{eq:#1})\xspace}%beginning of the sentence
\newcommand{\eqs}[2]{Eqs.\ (\ref{eq:#1},\ref{eq:#2})}

\newcommand{\tb}[1]{Table~\ref{tab:#1}\xspace}
\newcommand{\Tb}[1]{Table~\ref{tab:#1}\xspace}%beginning of the sentence
\newcommand{\se}[1]{Sect.~\ref{sec:#1}\xspace}
\newcommand{\Ap}[1]{Appendix~\ref{sec:#1}\xspace}

\newcommand{\tbr}[2]{\Phi_\mathrm{#1}^\mathrm{#2}}

\newcommand{\modeli}{\texttt{bC}\xspace}
\newcommand{\modelii}{\texttt{bCcgI}\xspace}
\newcommand{\modeliii}{\texttt{dCcgI}\xspace}
\newcommand{\modeliv}{\texttt{dCcgIcT}\xspace}
\newcommand{\modelv}{\texttt{dCcgIcL}\xspace}

\usepackage{ulem}
\makeatletter
\def\uwave{\bgroup \markoverwith{\lower3.5\p@\hbox{\sixly \textcolor{red}{\char58}}}\ULon}
\font\sixly=lasy6 % does not re-load if already loaded, so no memory problem.
\makeatother

\title[TRAPPIST-1: dynamical evolution]{The dynamics of the TRAPPIST-1 system in the context of its formation}

\author[Shuo Huang \& Chris W. Ormel]{
Shuo Huang (\begin{CJK*}{UTF8}{gbsn}黄硕\end{CJK*}),$^{1}$
Chris W. Ormel,$^{1}$\thanks{E-mail: chrisormel@tsinghua.edu.cn}
\\
$^{1}$Department of Astronomy, Tsinghua University, 100084 Beijing, China
}

\date{Accepted XXX. Received YYY; in original form ZZZ}

\pubyear{xxxx}

\begin{document}
\label{firstpage}
\pagerange{\pageref{firstpage}--\pageref{lastpage}}
\maketitle

% Abstract of the paper
\begin{abstract}
TRAPPIST-1 is an 0.09 $M_{\odot}$ star, which harbours a system of seven Earth-sized planets. Two main features stand out: (i) all planets have similar radii, masses, and compositions; and (ii) all planets are in resonance. Previous works have outlined a pebble-driven formation scenario where planets of similar composition form sequentially at the H$_2$O snowline (${\sim}0.1$ au for this low-mass star). It was hypothesized that the subsequent formation and migration led to the current resonant configuration. Here, we investigate whether the sequential planet formation model is indeed capable to produce the present-day resonant configuration, characterized by its two-body and three-body mean motion resonances structure. We carry out N-body simulations, accounting for type-I migration, stellar tidal damping, disc eccentricity-damping, and featuring a migration barrier located at the disc's inner edge. \changed{Due to the sequential migration, planets naturally form a chain of first-order resonances. But to explain the period ratios of the b/c/d-system, which are presently in higher-order resonances, we find that planets b and c must have marched across the migration barrier, into the gas-free cavity, before the disc has dispersed. We investigate both an early and late cavity infall scenario and find that the early infall model best matches the constraints, as well as being more probable.
After the dispersal of the gaseous disc, stellar tidal torque also contributes towards a modest separation of the inner system. We outline how the insights obtained in this work can be applied to aid the understanding of other compact resonant planet systems.}
\end{abstract}

% Select between one and six entries from the list of approved keywords.
% Don't make up new ones.

%%%%%%%%%%%%%%%%%%%%%%%%%%%%%%%%%%%%%%%%%%%%%%%%%%

%%%%%%%%%%%%%%%%% BODY OF PAPER %%%%%%%%%%%%%%%%%%

\section{Introduction}
TRAPPIST-1 is the first M8-dwarf star found to harbour seven transiting planets \citep{GillonEtal2016,GillonEtal2017}. While the orbits are located within 0.1\,au from the star in a compact (resonant) configuration, TRAPPIST-1's low luminosity renders planets d, e, and f to reside at equilibrium temperatures similar to Earth \citep{LugerEtal2017}. Therefore, the system is at the forefronts of characterization efforts and habitability studies \citep{LincowskiEtal2018} and a prime target of the James Webb Space Telescope \citep{GillonEtal2020,MeadowsEtal2021,LimEtal2021,LinEtal2021i}. A particular intriguing question is whether planet systems like TRAPPIST-1 are common. Lying within the 1,000 closest stars from the Sun, it is natural to conjecture that planet systems like TRAPPIST-1 must be common in the galaxy. 
%system popularly discussed especially on its occurrence among the universe. With non-detection of planet among a 44 brown dwarfs sample from Spitzer, \cite{HeEtal2017} compute the first limits on the presence of planets on close-in orbits and motivate future monitoring on more low mass stars.
However, \cite{SestovicDemory2020} and \cite{SagearEtal2020} argue that the seven planet TRAPPIST-1 discovery has been serendipitous, based on a study of over 500 low mass targets observed by the K2 spacecraft. Still, due to the challenging nature of detecting small planets around low-mass stars, the occurrence of TRAPPIST-1-like multi-planet systems is poorly constrained. 

To characterize the planets of the TRAPPIST-1 system, \cite{DelrezEtal2018} and \cite{GrimmEtal2018} evaluate the planetary radii and masses via transit depths and Transit Timing Variation (TTV) from Spitzer data. Recently, \cite{DucrotEtal2020} present all transits records for TRAPPIST-1 from Spitzer. \cite{AgolEtal2021} combine Spitzer observations \citep{DelrezEtal2018,DucrotEtal2020}, ground-based observations \citep{GillonEtal2016,GillonEtal2017,DucrotEtal2018,BurdanovEtal2019} and other space (K2 and HST) observations \citep{LugerEtal2017,GrimmEtal2018}, and refine the physical and dynamical properties of the TRAPPIST-1 planets. From these studies it follows that all seven planets in the TRAPPIST-1 system have roughly the same densities \citep{DornEtal2018,AgolEtal2021}, and similar low eccentricities ($e\lesssim0.01$) and inclinations ($i\lesssim0.01$) \citep{AgolEtal2021}. The resonant configuration is one of the most astonishing features in the TRAPPIST-1 system. From inner-to-outer, the period ratio of every adjacent planet pair is near 8:5, 5:3, 3:2, 3:2, 4:3, 3:2. The observed near-integer period ratios likely reflect corresponding Mean Motion Resonances (MMRs). In addition, every adjacent planet triplet is connected by various three-body resonances (3BRs) \citep{LugerEtal2017}.

\begin{figure*}
    \centering
    \includegraphics[width=1.8\columnwidth]{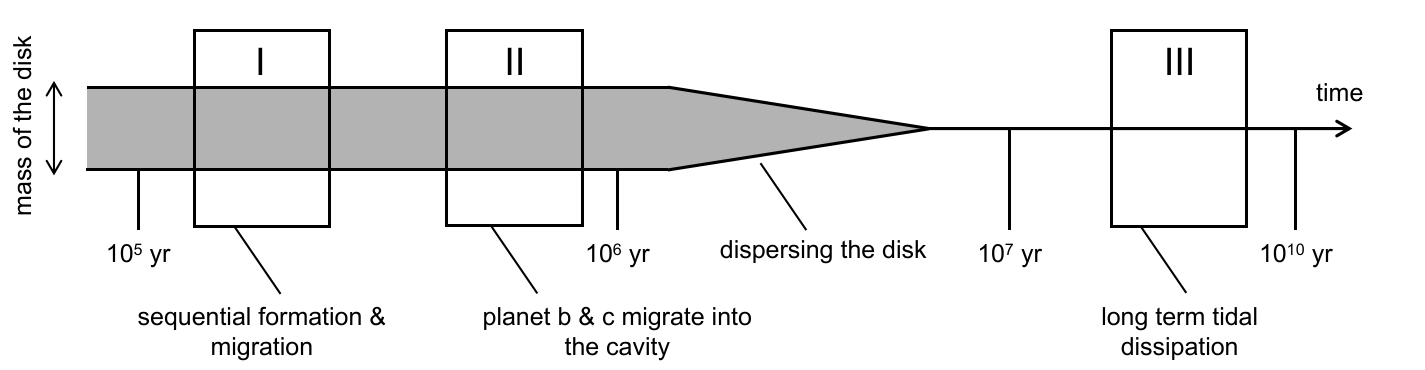}
    \caption{Schematic of the different stages in the dynamical evolution of the TRAPPIST-1 planet system. Time is oriented from  left to right with the mass of the disc indicated by the vertical width of the line. Different boxes indicate different events (stages) that occur over specific time intervals. In the Early Cavity Infall model, Stage I and II are contemporaneous. }
    \label{fig:old_model}
\end{figure*}

The similarity among the TRAPPIST-1 planets may reflect a similar formation origin. 
%Several formation mechanisms have been proposed to explain the composition similarities. 
Because of its compact nature, it is natural that the TRAPPIST-1 planets (or their building blocks) experienced significant radial migration. Planetesimals could originate within the snowline driven by the traffic jam effect of sublimated icy particles \citep{DrazkowskaEtal2016,IdaGuillot2016,SchoonenbergEtal2018}. 
On the other hand, re-condensation from outward diffusion of water vapour enhances the dust-to-gas ratio exterior the snowline, thus triggering streaming instability to form planetesimals \citep{SchoonenbergOrmel2017,OrmelEtal2017,HyodoEtal2019}. In both scenarios, planets are formed at the same location, which explains the common composition. Population synthesis models -- where planets are initialized at random locations -- also claim to reproduce the TRAPPIST-1-like properties \citep{BurnEtal2021}. After reaching a certain mass, planets migrate inwards, explaining the similar size of the planets, independent of whether growth is driven by planetesimal accretion or pebble accretion \citep{SchoonenbergEtal2019,ColemanEtal2019,BurnEtal2021,ZawadzkiEtal2021}. \changed{An intriguing question is how much water these planets were born with and how much still remains. According to  \citet{AgolEtal2021} the present-day water mass fraction are much lower than  prediction by \cite{OrmelEtal2017} and \cite{SchoonenbergEtal2019}.
However, $^{26}\mathrm{Al}$ isotropic heating can desiccate terrestrial planets \citep{LichtenbergEtal2019}. In addition, \cite{RaymondEtal2021} recently constrains that any water that (had been) present on the TRAPPIST-1 planets is likely to be accreted in the protoplanet disc phase.} 

%\ccc{rephrased and re-arranged some sentences}
Most of the above formation models did not address the dynamical evolution of the formed planets.  In the TRAPPIST-1 system, 3BRs bind planets together to render the planets long-term dynamically stable \citep{TamayoEtal2017}. At the same time, \cite{Mah2018} and \cite{BrasserEtal2019} find that the libration centre of the 3BR angles may jump to different values after tens of thousand years. Until now, only several planetary systems are known to harbour 3BRs: GJ 876 \citep{RiveraEtal2010}, Kepler-60 \citep{GozdziewskiEtal2016}, Kepler-80 \citep{Xie2013,MacDonaldEtal2016,MacDonaldEtal2021}, Kepler-223 \citep{MillsEtal2016}, K2-138 \citep{ChristiansenEtal2018,LopezEtal2019}, HD-158259 \citep{HaraEtal2020}, and TOI-178 \citep{LeleuEtal2021}. Recently, one 3BR is confirmed in Kepler-221 \citep{GoldbergBatygin2021} seven years after its discovery \citep{RoweEtal2014}. As hypothesized in this work, these resonances
%In particular, chains of planets in MMRs and 3BRs 
are fossils of the formation and evolution of planetary systems, which we can use to our advantage to reconstruct the planet formation history.
%may be leftover imprints of the planet formation process.
%If we able to trace the origin of these resonance configuration, we may be able to learn. Conceivably, We may find some clues left by planet formation when tracing back to the origin of these resonant configurations.\ccc{last sentence unclear...}

The convergent migration of planets in quasi-circular orbits generally results in first-order MMRs \citep{LeePeale2002,CorreiaEtal2018}. However, in the TRAPPIST-1 system, the innermost two planet pairs are near second or even third order MMRs. Conceivably, the innermost two planet pairs were near 3:2 MMR just after planet formation because it is the first order MMR closest to the current period ratio, whereafter the inner three planets separated from each other. \cite{LinEtal2021} presented an analytical model in which Earth-mass planets in discs around low mass stars end up in first order MMRs chain. However, they did not investigate the deviation from first order MMRs for the inner three planets. \cite{ColemanEtal2019}, using a population synthesis approach where planets form at arbitrary locations, observe several cases that have separated inner planets after planet formation and disc dispersal. However, the corresponding 3BRs are inconsistent with the observation. It is likely that a more specific, subtle mechanism operated to shape the dynamical structure of the inner planets. 

There are several ways to separate planets away from first-order MMR. In two-planets systems, divergent migration in the disc can drive planets away from MMRs, e.g., by magnetospheric rebound \citep{LiuEtal2017} or by adjusting the zero net torque location using a migration map approach \citep{WangEtal2021}. In the absence of a gaseous disc, stellar tidal dissipation can reduce the eccentricities of planets, and divergently migrate planets away from exact MMR \citep{LithwickWu2012,BatyginMorbidelli2013}. In multi-planet systems connected by 3BR, divergent migration, as well as tidal dissipation, can separate planets with each other maintaining 3BR relationships \citep{CharalambousEtal2018,PapaloizouEtal2018}. In particular, \cite{PapaloizouEtal2018} obtain the observed configuration of TRAPPIST-1 planets purely using stellar tidal dissipation. However, their initial conditions arbitrarily lie close to the present state, raising the question of how their initial state did materialize.
\changed{\cite{TeyssandierEtal2021} find that the observed high order resonances can be reproduced only with a smooth inner disc edge, but even then it is a rare simulation outcome.}
%\changed{\cite{TeyssandierEtal2021} adopt a smooth inner disc edge to explore the origin of the resonant chain of TRAPPIST-1, and can reproduce the observed TTVs, but they remark that the TRAPPIST-1 analog is a rare event \uwave{via only disc migration origination}. }

The goal of this paper is to connect the present long-term dynamical behaviour of the TRAPPIST-1 planets to their formation in the planet-forming disc. Our model consists of two steps.
In the first step, a chain of first-order MMRs is set up in the proto-planet disc. Planets sequentially form and migrate inward one after the other. Migration ceases at the disc's inner edge.
The second step involves the separation of the inner b/c/d subsystem until the observed period ratios are reached.
\changed{For this stage, we distinghuish between late infall, where planets b and c fall in the cavity after the first step has completed, and early infall, in which planets b and c cross the disc edge upon their arrival.}
After the disc disappears, long-term tidal dissipation further expands the planet system modestly, helping to finally achieve the observed resonant configuration. %\changed{According to model objectives, we divide our models into two categories: In the Late Infall model, we achieve the two steps separately. In the Early Infall model, we merge the two steps. }\ccc{this may be a bit confusing since you say there are two objectives and two steps.}

\begin{table*}
	\centering
	\caption{Radii, masses, semi-major axes and eccentricities of the TRAPPIST-1 planets. The values listed in the first three rows are used in the simulations. We compare the eccentricities in our simulation results to the eccentricities listed here from \citet{AgolEtal2021}. We also list the observed value of the 3BR angles in the TRAPPIST-1 system \citep{LugerEtal2017}. }
	\label{tab:data_agol}
	\begin{tabular}{lccccccc} % four columns, alignment for each
		\hline
		Planet: & b & c & d & e & f & g & h \\
		\hline
		$R[R_{\earth}]$ & 1.116 & 1.097 & 0.788 & 0.920 & 1.045 & 1.129 & 0.755 \\
        $M[M_{\earth}]$ & 1.374 & 1.308 & 0.388 & 0.692 & 1.039 & 1.321 & 0.326 \\
        $a[10^{-2}\,\mathrm{au}]$  & 1.154 & 1.580 & 2.227 & 2.925 & 3.849 & 4.683 & 6.189 \\
        $e$ & 0.004$\pm0.003^a$ & 0.002$\pm0.003$ & 0.006$\pm0.002$ & 0.0065$\pm0.0015$ & 0.009$\pm0.0015$ & 0.004$\pm0.0015$ & 0.0035$\pm0.0015$ \\
        3BR      &   &   $\tbr{b,c,d}{2,5,3} \approx 177^\circ$ & $\tbr{c,d,e}{1,3,2}\approx49^\circ$ &   $\tbr{d,e,f}{2,5,3}\approx-148^\circ$ & $\tbr{e,f,g}{1,3,2}\approx-76^\circ$ & $\tbr{f,g,h}{1,2,1}\approx177^\circ$ \\
		\hline
		\multicolumn{6}{l}{$^a$ The error bars are estimated from Fig. 21 in \citet{AgolEtal2021}. }
	\end{tabular}
\end{table*}

\begin{table}
	\centering
	\caption{Definition of quantities, symbols and units. The last five parameters separated by horizontal line are the free parameters of the simulations. }
	\label{tab:params}
	\begin{tabular}{p{0.25cm}p{3cm}ll} % four columns, alignment for each
		\hline
		 & Description & Value & Reference \\
		\hline
		$h$ & aspect ratio & 0.03 & \se{disc}\\
		$\dot{M}_{\mathrm{g}}$ & star accretion rate & $10^{-10}M_{\sun}\mathrm{yr^{-1}}$ & \se{disc} \\
		$M_{\star}$ & stellar mass & 0.09$M_{\sun}$ & \se{disc} \\
		$\gamma_{I}$ &  in type-I migration rate & 2 & \eqs{tm}{ta_n} \\
		$\tau_{\mathrm{d}}$ & disc dissipation time-scale & $10^{5} \,\mathrm{yr}$ & \se{numerical} \\ 
		$\Delta$ & migration threshold width & $2h r_{\mathrm{c}}$ & \eqs{fa}{fe} \\
		$Q_\mathrm{sim}$ & tidal dissipation parameter & 100(0.1 in Stage~III) & \eq{tidal} \\
		\hline
		$\tau_{a,\earth}$ & nominal semi-major axis damping timescale & [$1-40\,\mathrm{kyr}$] & \eq{ta_n} \\
		$C_e$ & eccentricity-damping prefactor & [$0.1-1$] & \eq{te_d} \\
		$r_{\mathrm{c}}$ & truncation radius & [$0.013-0.030\,\mathrm{au}$] & \eqs{fa}{fe} \\
		$A_a$ & migration threshold height & 50, 100, 150 & \eq{fa} \\
		$A_e$ & enhancement of disc eccentricity-damping & $[1-40]$ & \eq{fe} \\
		
		\hline
	\end{tabular}
\end{table}

Using N-body simulations, we first examine several simple models that fail to establish a more separated inner subsystem. We conclude that planets b and c had to migrate interior to the disc truncation radius, which allowed them to separate already during the disc phase. Finally, accounting for long-term stellar tidal damping, we manage to reproduce the dynamical configuration of TRAPPIST-1 planets including their period ratios, eccentricities, and 3BRs.

\changed{
This paper is structured as follows. In \se{model}, we describe the terminologies and our physical method. We then present the results from simple models and assess the conditions for meeting Objective I in \se{simp_model}. Turning to Objective II, we carry out a parameter study to search for the conditions that can successfully reproduce our objectives (in \se{3BR_test}). We then examine the Early Infall model in \se{new_model}. Analytical supports for our findings is given in \se{analysis}. Finally, we assess our models and the significance of our findings in \se{discussion} and conclude in \se{conclusion}. }

\section{Model}
\label{sec:model}
\subsection{Preliminaries and Terminology}
\subsubsection{Simulation stages}
\label{sec:stages}
We conduct N-body simulations to integrate the planets equation of motion. In this work we follow the dynamical evolution of planets in three stages as shown in \fg{old_model}:
\begin{enumerate}
    \renewcommand{\labelenumi}{(\Roman{enumi})}
    \item the stage after their formation, where planets migrates inwards in the gas disc and end up in a chain of first-order resonances;
    \item the stage where planets b and c enter the gas-free cavity;
    \item the post-disc stage where stellar tides operate on $\sim$Gyr time-scales.
\end{enumerate}
A key simplification is to simulate each of these phases separately, instead of running one simulation that contains all these stages. Not every model contains each of these three stages. \changed{Usually Stage~I, II and III occur in chronological order, but in the Early Infall model in \se{new_model}, Stage~I and Stage~II happen simultanously. }
%\ccc{maybe we dont need to mention the models by name}
%\modeli and \modelii feature Stage~I and~III. \modeliii feature Stage~I, Stage~II and Stage~III. \modeliv features all of the three stages.

\subsubsection{Parameters}
All masses, semi-major axes and radii of the TRAPPIST-1 planets are adopted from \cite{AgolEtal2021} and are listed in \tb{data_agol}. The values of the default parameters and the range of free parameters appearing in the equations of this paper are listed in \tb{params}.

\subsubsection{Resonance angles}
For two planets $i$ and $i+1$, a two-body $j-o_{i}{:}j$ MMR angle can be defined
\begin{equation}
    \phi_{i,i+1;i+X} = (j-o_{i}) \lambda_i -j\lambda_{i+1} +o_{i}\varpi_{i+X}
\end{equation}
where $o_{i}$ is the order of the resonance, $\lambda_i$ and $\lambda_{i+1}$ are the mean longitudes of planets $i$ and $i+1$, and $\varpi_{i+X}$ is either the longitude of periapsis of planet $i$ ($X=0$) or $i+1$ ($X=1$). A similar MMR angle can be defined for the $(k-o_{i+1}){:}k$ MMR of planets $i+1$ and $i+2$. Then we can define the 3BR angle such that it does not involve the longitudes of pericentre:
\begin{equation}
    \Phi_{i,i+1,i+2}^{p,p+q,q}
%    = o_{i+1}\phi_{i,i+1;i+1} -o_i\phi_{i+1,i+2;i+1}
    = p\lambda_{i} -(p+q)\lambda_{i+1} +q\lambda_{i+2}
\end{equation}
where $p=o_{i+1}(j-o_{i})$ and $q=o_{i}k$. Observationally, it is hard to constrain the longitudes of pericentre when, as is typical, the eccentricity is low. The three-body angles can however be constrained more easily. TRAPPIST-1's c/d/e planets are currently near 3:5 and 2:3 MMRs, corresponding to $p=3$ and $q=6$, and the 3BR angle $\tbr{c,d,e}{3,9,6} = 3\tbr{c,d,e}{1,3,2}$. We list the five main 3BR angles of the TRAPPIST-1 system in \Tb{data_agol}.

\subsection{Numerical integration}
\label{sec:numerical}
Every planet obeys the equation of motion which includes the two body forces from the star and other planets, the angular momentum loss and orbital circularization induced by the gas disc and damping from tides with the star:
\begin{equation}
  \boldsymbol{F}_{\mathrm{i}}=\sum_{j\neq i}^N \frac{Gm_j}{\left|r_{ji}\right|^3}\boldsymbol{r}_{\mathrm{ji}}+\boldsymbol{F}_{\mathrm{disc},i}+\boldsymbol{F}_{\mathrm{star},i}.
  \label{eq:eom}
\end{equation}
In \eq{eom} the first term on the right-hand side is the standard gravitational interaction N-body force, where $\bm{r}_{ji} = \bm{r}_j -\bm{r}_i$ and the index $j$ represent the planets other than planet $i$. The term $\boldsymbol{F}_{\mathrm{disc},i}$ indicates the gas disc tidal force and $\boldsymbol{F}_{\mathrm{star},i}$ indicates the stellar tidal damping force acting on planet $i$.

Simulations were conducted using the \verb'WHFast' integration method in \verb'REBOUND' \citep{ReinLiu2012} and our additional forces were  implemented via \verb'REBOUNDx' \citep{TamayoEtal2020}. In \eq{eom}, only the $\boldsymbol{F}_{\mathrm{disc}}$ term is proportional to the disc gas density.  During the disc dispersing, we add an exponential decay factor $\exp[{-(t-t_0)/\tau_{\mathrm{d}}}]$ to $\boldsymbol{F}_{\mathrm{disc}}$. In addition, the disc dispersal time-scale $\tau_{\mathrm{d}}$ is taken to be $10^5$ years. We ignore the disc torque term in the post-disc environment of stage III. The time-step during the simulations is less than 5\% of the orbital period of the innermost planet.

\subsection{Planet-disc interaction}
\label{sec:disc}
The planet-disc interaction can cause planet migration \citep{GoldreichTremaine1980, Ward1997} and eccentricity-damping \citep{Ward1988, Artymowicz1993, TanakaWard2004}. Following \citet{PapaloizouLarwood2000} we write the disc tidal force as:
\begin{equation}
  \boldsymbol{F}_{\mathrm{disc},i}=-\frac{\boldsymbol{v}_{\mathrm{i}}}{2\tau_{\mathrm{a},i}}-\frac{2(\boldsymbol{v}_{\mathrm{i}}\cdot\boldsymbol{r}_{\mathrm{i}})\boldsymbol{r}_{\mathrm{i}}}{\left|r_i\right|^2\tau_{\mathrm{e},i}^{\mathrm{d}}}
  \label{eq:fdisc}
\end{equation}
where $\tau_{\mathrm{a},i}$ and $\tau_{\mathrm{e},i}^{\mathrm{d}}$ indicate the semi-major axis and eccentricity-damping time-scales of planet $i$ due to migration. 

We assume the gas disc is truncated at a radius $r_{\mathrm{c}}$ by the magnetic field from the star \citep{PringleRees1972,D'AngeloSpruit2010}. We assume the inner disc is viscously relaxed \citep{ShakuraSunyaev1973, Lynden-BellPringle1974} within the snowline $r_{\mathrm{ice}}\sim0.1\,\mathrm{au}$ \citep{SchoonenbergEtal2019, LinEtal2021}, implying constant $\dot{M}_g=3\pi\nu\Sigma$ where $\nu$ is the viscosity \changed{and $\dot{M}_g$ is taken to be $10^{-10}M_{\sun}\mathrm{yr^{-1}}$ \citep{ManaraEtal2015}.} The aspect ratio is fixed at $h=0.03$ following \cite{OrmelEtal2017}, whose value is motivated by viscous heating and lamppost heating \citep{RafikovDeColle2006} and the spectral energy distribution fitting \citep{MuldersDominik2012}. In an alpha-disc model, $\nu=\alpha h^2 \Omega r^2$ with constant $\alpha$ parameter \citep{ShakuraSunyaev1973}, the disc surface density is therefore: 
\begin{equation}
    \Sigma_{\mathrm{g}} = \Sigma_0 \left(\frac{r}{r_0}\right)^{-0.5}
\end{equation}
where $\Sigma_0$ is the disc surface density at distance $r_0$ ($< r_{\mathrm{ice}}$).  Planets' semi-major axes and eccentricities are damped on time-scales:
\begin{equation}
    \label{eq:tm}
    \tau_{\mathrm{a},i} = \frac{h^2/f_{\mathrm{a}}}{2\gamma_I q_{{i}} q_{\mathrm{gas}}\Omega_{\mathrm{K}}} = \tau_{\mathrm{a,\earth}} \left(\frac{q_{i}}{q_{\earth}}\right)^{-1} \left(\frac{1}{f_{\mathrm{a}}}\right)
    %\frac{3\pi \alpha h^4}{\gamma_{I} q_{pl} {\dot{M}_g} /M_{\star}}
\end{equation}
\begin{equation}
    \tau_{\mathrm{e},i}=C_eh^2\tau_{\mathrm{a},i}\frac{f_{\mathrm{a}}}{f_{\mathrm{e}}},
    \label{eq:te_d}
\end{equation}
where $\gamma_I$ is the prefactor in the type~I migration torque expression and $q_{i}=m_{i}/M_{\star}$ \citep{TanakaEtal2002,KleyNelson2012}. The semi-major axis damping time-scale is half of the angular momentum damping time-scale \citep{TeyssandierTerquem2014}. $\tau_{\mathrm{a,\earth}}$ is the nominal Type~I semi-major axis damping time-scale:
\begin{align}
    \tau_{\mathrm{a,\earth}}
    &= \frac{h^2}{2 \gamma_I q_{{\earth}} q_{\mathrm{gas}}\Omega_{\mathrm{K}}} \\
    &= 1.2\times 10^4 \left( \frac{\gamma_I}{2}\right)^{-1}\left(\frac{\dot{M}_g}{10^{-10}M_{\sun}\mathrm{yr^{-1}}}\right)\left(\frac{\alpha}{10^{-3}}\right) \, \mathrm{yr},
    \label{eq:ta_n}
\end{align}
where $q_{{\earth}}$ stand for the mass ratio of an Earth mass planet over the stellar mass and $q_{\mathrm{gas}}=\Sigma_{\mathrm{g}}r^2/M_{\star}$. 
% \corem{ Since $\alpha$ is the only free parameter in \eq{ta_n}, we use $\tau_{\mathrm{a,\earth}}$ directly during N-body integration for simplicity}. \ccc{rather cryptic sentence, perhaps: (or we can delete)} 
\changed{Therefore $\tau_{a,\earth}$ and $C_e$ are the parameters that control the damping of the planets in the N-body simulations.}

\changed{
In \eq{te_d}, $C_e$ expresses the rate of eccentricity-damping mechanism over semi-major axis damping. In the linear theory its value is 1.28 \citep{TanakaWard2004}, which is consistent with the 3D hydro-dynamical simulations \citep{CresswellNelson2008}. 
However, \cite{CresswellNelson2006} suggest that $C_e\sim0.1$ is sometimes needed to fit their hydro-dynamical simulations while generally this parameter encapsulates the uncertainty in (and variety of) disc migration mechanisms, which amount to considerable uncertainty in the type-I migration prefactor $\gamma_I$ \citep{PaardekooperEtal2011, BaruteauEtal2014,Benitez-LlambayEtal2015,McNallyEtal2017}.} \changedd{When the disc surface density gradient ($-\mathrm{d}\ln{\Sigma_g}/\mathrm{d}\ln{r}$) and temperature gradient ($-\mathrm{d}\ln{T}/\mathrm{d}\ln{r}$) become small, Type~I migration slows down ($\gamma_I$ decreases) and $C_e$ goes down.} 
\changed{In this study we take an agnostic view and examine $C_e$ values in the range of 0.05 to 1.}

\begin{figure}
	% To include a figure from a file named example.*
	% Allowable file formats are eps or ps if compiling using latex
	% or pdf, png, jpg if compiling using pdflatex
	\centering
	\includegraphics[width=\columnwidth]{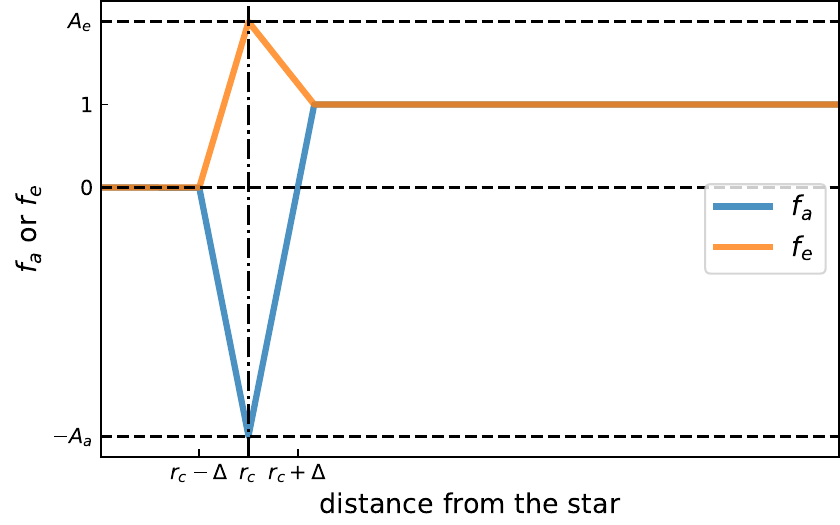}
    \caption{Sketch of $f_{\mathrm{a}}$ and $f_{\mathrm{e}}$. The $x$-axis denotes the distance measured from $r_{\mathrm{c}}$, the $y$-axis the value of $f_{\mathrm{a}}$ or $f_{\mathrm{e}}$. The horizontal black-dashed lines indicate $A_a$ and $A_e$. The vertical dashed dotted line represents the location of $r_{\mathrm{c}}$.}
    \label{fig:torque_sketch}
\end{figure}

\begin{table*}
    \centering
    \caption{Summary of our model designs. }
    \label{tab:model}
    \begin{tabular}{|l|ccc|ccc|cc|p{6.5cm}}
    \hline
    Model suite     & \multicolumn{3}{c}{Properties $^a$} & \multicolumn{3}{c}{Stages included$^b$} & \multicolumn{2}{c}{Objective$^c$} & Assessment \\
                    & $r_\mathrm{c}$ [au] & Formation & Disc  & I & II & III & I & II \\ 
                    & & imprint & repulsion \\
    \hline          
    \modeli   & 0.013 &    -    & N & \checkmark &   $\times$ & \checkmark & $\times$   & $\times$ & cannot achieve all first order resonances  \\
    \modelii  & 0.013 & c \& g  & N & \checkmark &   $\times$ & \checkmark & \checkmark & $\times$ & period ratio outer planets expand too much \\
    \modeliii & 0.023 & c \& g  & N & \checkmark & \checkmark & \checkmark & \checkmark & $\times$ & stellar tidal force alone cannot break 3BR of planet c\\
    \modeliv  & 0.023 & c \& g  & Y & \checkmark & \checkmark & \checkmark & \checkmark & \checkmark & success with inward torque on c and high $e$-damping on d \\
    \modelv   & 0.023 & c \& g  & Y & \checkmark & \checkmark & \checkmark & \checkmark & \checkmark & success with early infall of b and c in cavity; more likely  \\
    \hline
    % {$^a$ Given are the cavity radius (disc inner edge) $r_\mathrm{c}$, which planets are considered as the planet formation imprint, and the presence of a disc repulsion. }\\
    % \multicolumn{10}{l}{$^b$ Stages with \checkmark are included in a model while ones with $\times$ are excluded. }
	\end{tabular}
	\flushleft
	{$^a$ Listed are the cavity radius (location of disc inner edge) $r_\mathrm{c}$, which planets are considered as the planet formation imprint, and the presence of disc repulsion on planets in the cavity. }\\
	{$^b$ Stages with \checkmark are included in a model while ones with $\times$ are excluded. Different stages are introduced in \fg{old_model}. }\\
	{$^c$ The model that meets the objective is marked by \checkmark, if not we place $\times$ instead. The two objectives are introduced at the beginning of \se{simp_model}.}
\end{table*}

The parameters $f_{\mathrm{a}}$ and $f_{\mathrm{e}}$ determine the behaviour of semi-major axis and eccentricity-damping near the disc inner edge. When the location of one planet $r\gg r_{\mathrm{c}}$, $f_{\mathrm{a}}=1$. If the planet reaches the disc's inner edge, only Lindblad torques from the exterior disc remain. Similarly, only the upper horseshoe motion and associated co-orbital torque is present. In other words, the torques have become one-sided. For low-mass planets, the (positive) one-sided co-rotation torque will win out over the (negative) Lindblad torques, halting migration \citep{LinPapaloizou1979,LinPapaloizou1993,PaardekooperPapaloizou2009,LiuEtal2017}. Therefore, $f_{\mathrm{a}}$ must change sign (become negative) at the disc's inner edge. We adopt an expression for $f_{\mathrm{a}}$ in \eq{tm} as:
\begin{equation}
    f_{\mathrm{a}}=\left\{\begin{array}{ll}
        0 & (r<r_{\mathrm{c}}-\Delta)\\
        A_a\frac{r_{\mathrm{c}}-\Delta-r}{\Delta} & (r_{\mathrm{c}}-\Delta<r<r_{\mathrm{c}}) \\
        A_a\frac{r-r_{\mathrm{c}}-\Delta}{\Delta} & (r_{\mathrm{c}}<r<r_{\mathrm{c}}+\Delta+1/A_a) \\
        1 & (r>r_{\mathrm{c}}+\Delta+1/A_a)
    \end{array}
    \right.
    \label{eq:fa}
\end{equation}
where $\Delta$ and $A_a$ determine the width and strength of the trap. If sharply truncated at the disc's inner edge, $\Delta$ is comparable to the half-width of the horseshoe region so $\Delta\sim H_\mathrm{c}$. For simplicity, we take $\Delta=2H_\mathrm{c}$, where $H_\mathrm{c}$ is the scale height of gas disc at truncation radius. \changed{According to \cite{LiuEtal2017}, $A_a$ follows as the ratio of the (positive) one-side corotation torque \citep[eq. 11 in][]{LiuEtal2017} over the Type~I migration torque (two times the reciprocal of \eq{tm} times angular momentum)}
\begin{equation}
    A_a = \frac{\Gamma_\mathrm{c,1s}}{\Gamma_\mathrm{I}}
    = \frac{C_\mathrm{hs}}{\gamma_I h} \left( \frac{q_p}{h^3} \right)^{-1/2}
    = \frac{135}{\gamma_I} \left( \frac{q_p}{10^{-5}} \right)^{-1/2} \left( \frac{h}{0.03}\right)^{1/2}
\end{equation}
where $C_\mathrm{hs}=2.46$.
\changed{For a planet-to-star mass ratio of $q=10^{-5}$ and $h=0.03$ this gives $A_a=68$. Therefore we take $A_a$ around this value.}
\Eq{fa} is similar in shape to the "deep drop" model \cite{AtaieeKley2021} adopt in their N-body simulations.

In our simulations, we find that it is sometimes advantageous to enhance the eccentricity damping of planets at the disc edge ($f_e>1$). Eccentricity damping of planets could be more efficient near the disc's inner edge for the following reasons:
\begin{itemize}
    \item Material may pile up significantly near $r_{\mathrm{c}}$ if $r_{\mathrm{c}}$ is larger than the corotation radius of the star \citep{D'AngeloSpruit2010};
    \item The absence of inner Lindblad resonances (they lie in the cavity), which would excite planets' eccentricity \citep{Ward1988};
    \item hydro-dynamical simulation indicate that eccentrities for planets at the disc inner edge are more strongly damped \citep[see Fig.15 of][]{AtaieeKley2021}.
\end{itemize}
Therefore we express $f_{\mathrm{e}}$ in \eq{fdisc} with a similar expression as $f_\mathrm{a}$:
\begin{equation}
    f_{\mathrm{e}}=\left\{\begin{array}{ll}
        0    & (r<r_{\mathrm{c}}-\Delta)\\
        A_e\frac{r-r_{\mathrm{c}}+\Delta}{\Delta} & (r_{\mathrm{c}}-\Delta<r<r_{\mathrm{c}}) \\
        (A_e-1)\frac{r_{\mathrm{c}}+\Delta+1/A_a-r}{\Delta+1/A_a}+1 & (r_{\mathrm{c}}<r<r_{\mathrm{c}}+\Delta+1/A_a) \\
        1 & (r>r_{\mathrm{c}}+\Delta+1/A_a)
    \end{array}
    \right.
    \label{eq:fe}
\end{equation}
where $A_e$ is the enhancement of the eccentricity-damping near the truncation radius. Not every simulations feature this anomalous eccentricity damping at the disc edge.
% \changed{In the improved Late Infall model (\se{3BR_test}), we vary $A_e$ from 1 to 40, although very high values may be unfeasible. }

We plot $f_{\mathrm{a}}$ and $f_{\mathrm{e}}$ in \fg{torque_sketch}. Near the truncation radius, $f_{\mathrm{a}}$ decreases sharply and becomes negative, thus pushing planets in this region outward. The outward torque peaks at $r_{\mathrm{c}}$. Within $r_\mathrm{c}$, the outward migration torque decreases to zero because there is no gas in the disc cavity (See \se{oneside}, however, for the effects of a distant disc torque on planets inside the cavity).

\subsection{Tidal dissipation}
For a compact planet system in which planets are close to their host star, such as in TRAPPIST-1, tidal dissipation can be significant. The stellar tidal damping force term in \eq{eom} is:
\begin{equation}
    \label{eq:tidal_force}
    \boldsymbol{F}_{\mathrm{star},i} = -\frac{2(\boldsymbol{v}_{\mathrm{i}}\cdot\boldsymbol{r}_{\mathrm{i}})\boldsymbol{r}_{\mathrm{i}}}{\left|r_i\right|^2\tau_{\mathrm{e},i}^{\mathrm{s}}},
\end{equation}
where $\tau_{\mathrm{e},i}^{\mathrm{s}}$ refers to the eccentricity-damping time-scale induced by tidal interaction with the host star:
\begin{equation}
    \tau_{\mathrm{e},i}^{\mathrm{s}}=7.6\times10^5Q_{\mathrm{sim}}\left(\frac{m_i}{m_{\earth}}\right)\left(\frac{M_{\sun}}{M_{\star}}\right)^{1.5}\left(\frac{R_{\earth}}{R_i}\right)^5\left(\frac{a_i}{0.05\, \mathrm{au}}\right)^{6.5} \,\mathrm{yr},
    \label{eq:tidal}
\end{equation}
where $Q_{\mathrm{sim}}=3Q/(2k_2)$, $Q$ is the tidal dissipation function and $k_2$ is the Love number \citep{GoldreichSoter1966,PapaloizouEtal2018}. For solar system planets in the terrestrial mass range, $Q_{\mathrm{sim}}$ is estimated between 50--2500. In this paper we take $Q_{\mathrm{sim}}=100$ in most cases, similar to the terrestrial planets in our solar system. 
In Stage~III, following \citet{PapaloizouEtal2018}, we take $Q_{\mathrm{sim}}=0.1$ to accelerate the simulation by a factor 1,000.
%When we run long term tidal dissipation simulations without disc (in Stage~III), we take for the innermost two planets b \& c to accelerate the simulation by one thousand times. This approach was used by \cite{PapaloizouEtal2018} as well.

\section{Preliminary simulations}
\label{sec:simp_model}
%\outline{Section design. Tell the models features and which section every model is located at. Show \tb{model} as an instruction of all models, in which, for example model 1 is explanined as below. Truncation radius mode: close, i.e., 0.013au. Initial location of planet g mode: outside 3:2 or 2:1 (they are equivalent). The torque from one side disc push on planets in the cavity mode: no. The model comparison is in \se{comparison}.} 
In this section, we summarize our model design and the principal outcomes. This work concentrates on the dynamic evolution after the planet formation process. We insert planets sequentially at the beginning of the simulation with fixed masses following \tb{data_agol}. \cite{OrmelEtal2017}, \cite{SchoonenbergEtal2019} and \cite{LinEtal2021} hypothesize that the TRAPPIST-1 planets are formed at the snowline (${\approx}0.1$\,au) and quickly accreted material\changed{, while \citet{ColemanEtal2019} and \citet{BurnEtal2021} initialize many planetary embryos in the disc to grow them to TRAPPIST-1 planets}. In the context of this work planets could also have formed further out; but we simply fix the mass of every planets during the simulation.

Due to type~I migration \citep{TanakaEtal2002}, planets migrate towards the host star when they grow massive enough and stop at the disc truncation radius. 
The migration speed is so fast that planets tend to be trapped in first order MMR while avoiding the higher-order MMR. Therefore, we initialize every new planet just beyond the 2:1 MMR location of the outermost planet by default. Still, by virtue of the short formation time \citep{LinEtal2021} planets may be born near 3:2 and 4:3 period ratios with the inner planet. Therefore, it is possible that planets are born at locations within the 3:2 MMR. We investigate whether this imprint of planet formation is needed to reproduce the orbital configuration. 

Therefore, our first step is to park all planets in a chain of first order MMR, i.e., 3:2, 3:2, 3:2, 3:2, 4:3, 3:2 (hereafter referred to as Objective~I). However, this contrasts the present-day dynamical state where the period ratio of planets c/b and d/c are 8/5 and 5/3, respectively, i.e., close to higher-order MMRs. Our second step is to evolve the first order MMR chain to the 8:5, 5:3, 3:2, 3:2, 4:3, 3:2 period ratios chain (hereafter Objective~II). We separate the simulation into three stages as introduced in \fg{old_model}.

\begin{figure}
	% To include a figure from a file named example.*
	% Allowable file formats are eps or ps if compiling using latex
	% or pdf, png, jpg if compiling using pdflatex
	\centering
	\includegraphics[width=\columnwidth]{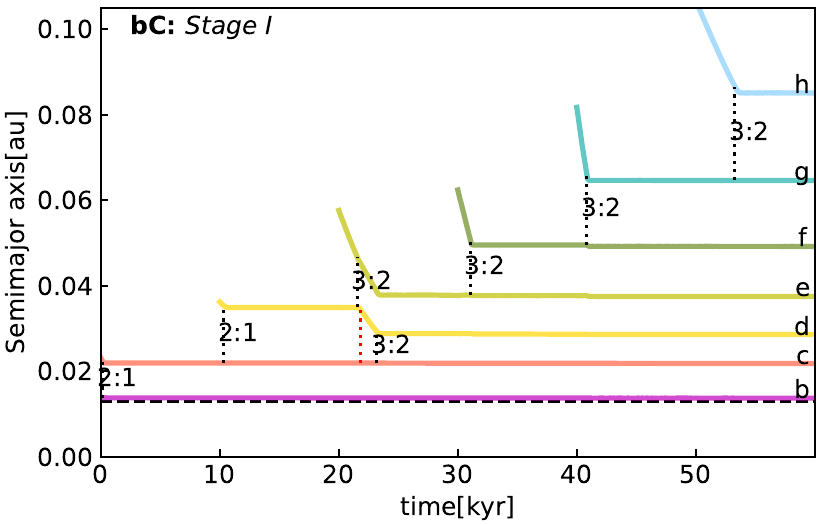}
    \caption{One-by-one migration of the TRAPPIST-1 planets after their sequential formation in the proto-planetary disc for a nominal migration time of $\tau_{\mathrm{a,\earth}}=5\times10^3\,\mathrm{yr}$ and $C_e=0.1$. The eccentricity enhancement at the disc inner edge is $A_e=1$. Solid lines with different colours indicate the semi-major axes of different planets. The horizontal black dashed line indicates the position of the migration barrier $r_{\mathrm{c}}$. Vertical black and red dashed segments between every two planets trajectories indicate the moment planets converge to and depart from 2:1 and 3:2 period ratios. }
    \label{fig:model1_sma}
\end{figure}

The set of our models is listed in \Tb{model}. Different columns express model characteristics. In \modeli model, we set the cavity radius at $r_{\mathrm{c}}=0.013\,\mathrm{au}$, which coincides with the observed location of planet b. In the \modelii model, we insert planet g interior to the 3:2 MMR location. In the \modeliii model, we set $r_{\mathrm{c}}=0.023\,\mathrm{au}$, which coincides with the observed location of planet d, and insert planet g interior to 3:2. In the \modeliv model, we additionally switch on the one-sided repulsive torque from the gas disc on planet c in the cavity\changed{, while in \modelv we use a particular choice for this repulsive force}. The first three models will be discussed in this section. Model \modeliv and \modelv will be discussed in \se{3BR_test} and \se{new_model}, respectively.

\subsection{Planet b at the cavity edge (\modeli)}
\label{sec:model1}
Objective~I -- the parking of planets in a chain of first order MMRs close to the observed period ratios -- can most straightforwardly be achieved by the planet's sequential migration and pileup near $r_c$. Therefore, we let $r_{\mathrm{c}}$ coincide with the location of planet b. We vary $\tau_{\mathrm{a,\earth}}$ and $C_e$, and run a set of Stage~I simulations;  $A_e$ is taken to be unity. In the simulations, we insert a new planet every ${\sim}10{\,\mathrm{kyr}}$ (long enough for the inner planets to stabilize). For simplicity, We initialize planet b near $r_{\mathrm{c}}$, and then the other planets outside the 2:1 period ratio.

One of our typical outcomes is shown in \fg{model1_sma}. We adopt $\tau_{\mathrm{a,\earth}}=5\times10^3\,\mathrm{yr}$ and $C_e=0.1$. Planet c gets trapped by the 2:1 MMR with planet b. Similarly, planet d gets trapped by the 2:1 MMR with planet c. Planet e directly crosses the 2:1 MMR because planets d \& e are less massive and further from the star than planet b, c, and d. Therefore, the Type-I migration can overcome the resonant repulsion and it gets trapped by the 3:2 MMR with planet d. At the same time, planet d crosses the 2:1 MMR because of the pushing from resonant interaction with planet e. Afterwards, the later planets are trapped by the 3:2 MMR with their former peers. The truncation radius acts to prevent planets migrating into the cavity. At the end of this simulation, the period ratios of every two adjacent planets are about 3:2, except planets b \& c, which is about 2:1. However, Objective~I requires planets b \& c to park in 3:2 and planets f \& g in 4:3. 

\begin{figure}
    \centering
    \includegraphics[width=\columnwidth]{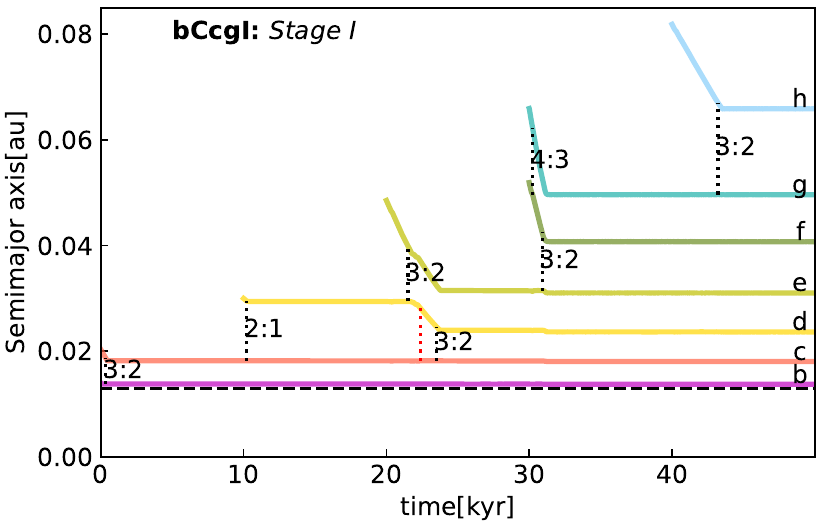}
    \caption{Similar to \fg{model1_sma}, but with planet c starting out inside the 2:1 period ratio and planet g initially inside the 3:2 period ratio. We take $\tau_{\mathrm{a,\earth}}=5\times10^3\,\mathrm{yr}$, $C_e=0.1$ and $A_e=1$.}
    \label{fig:model2_sma}
\end{figure}

\begin{figure*}
    \centering
    \includegraphics[width=1.6\columnwidth]{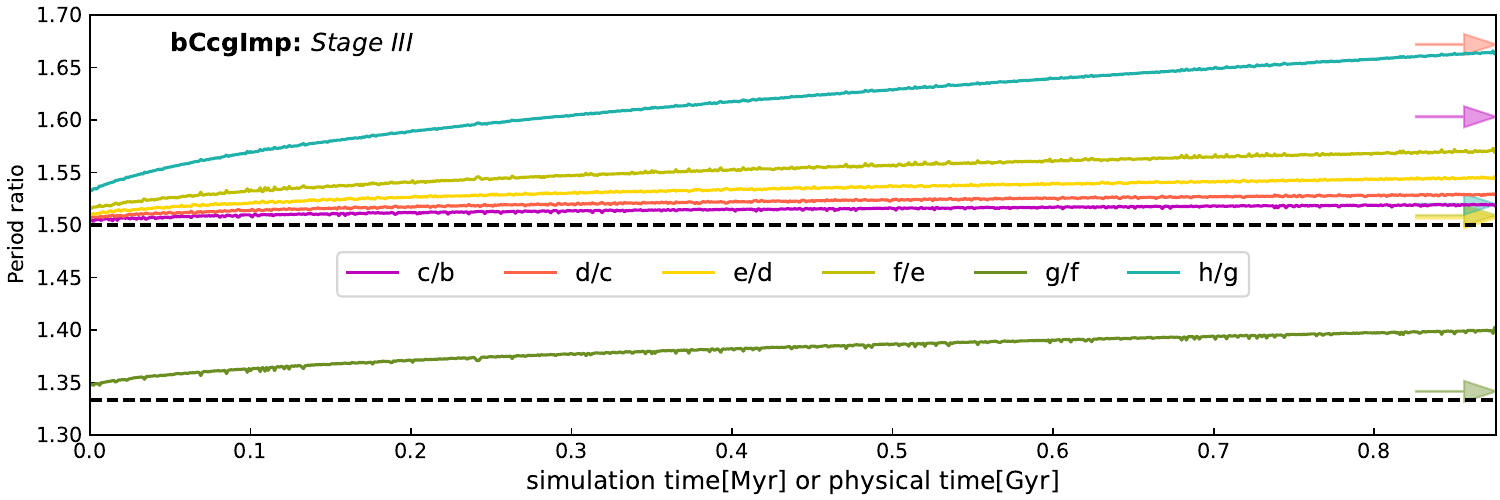}
    \includegraphics[width=0.8\columnwidth]{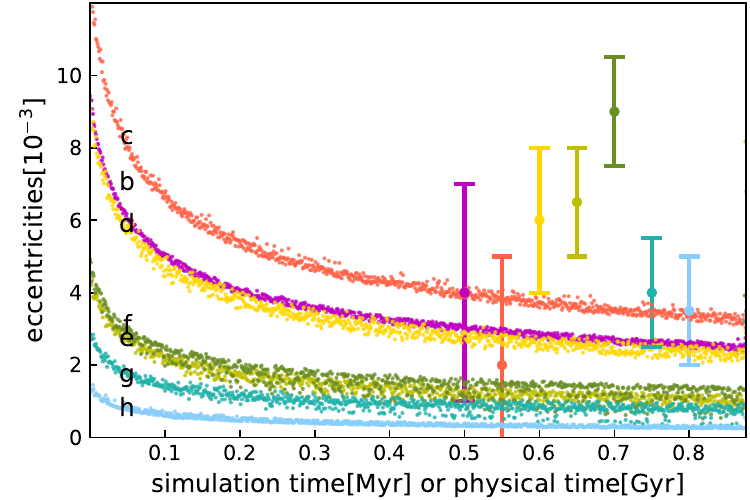}
    \includegraphics[width=0.8\columnwidth]{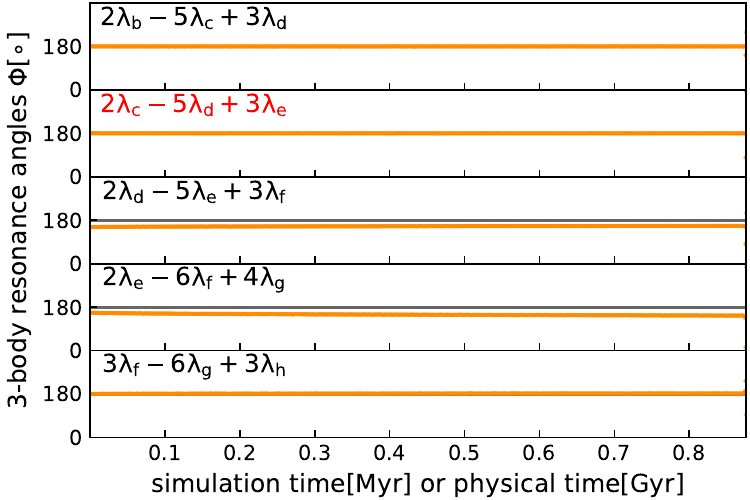}
    \caption{Period ratios (top panel), eccentricities (bottom-left panel) and 3BR angles (bottom-right panel) during tidal dissipation (with $Q_{\mathrm{sim}}=0.1$). In the period ratios plot, dashed lines indicate the 3:2 and 4:3 period ratios. Arrows indicate the observed value of the period ratios. \changed{In the eccentricity plot, we label the name of each planets on the left and indicate the present-day eccentricities by error bars with different colours (the x-coordinate of the error bars does not have meaning).} In the 3BR angles plots, we label the expression of each angle on the upper left of each panel. The angle labelled in red does not librate at present. }
    \label{fig:model2_separate}
\end{figure*}

Naively, closer-spaced MMR can be obtained with a faster migration rate (lower $\tau_{a,\earth}$). However, if we take $\tau_{\mathrm{a,\earth}}<5\times10^3\, \mathrm{yr}$ to migrate planets more rapidly, it results in the breaking of the 3:2 MMR of planet d \& e. Therefore, the value of the nominal migration time-scale $\tau_{\mathrm{a,\earth}}$ must exceed $5\times10^3\,\mathrm{yr}$ when the cavity radius is fixed at $r_{\mathrm{c}}=0.013\,\mathrm{au}$. Consequently, either b/c and f/g end up in too wide MMR or d/e ends up in a too close MMR, thus failing to accomplish Objective~I. 

\subsection{Formation imprints for planet c and g (\modelii)}
\label{sec:model2}
Instead of accelerating planet migration, the closer MMR of planet b \& c and planet f \& g can be obtained by reducing the time/space span between their formation -- a condition that we refer to an imprint of planet formation. In the \modelii model, the parameters in the simulation are the same as in the \modeli model. We make the additional assumption that planets c and g are born within 2:1 and 3:2 period ratios in Stage~I. One of our typical outcomes is shown in \fg{model2_sma}. Planet c gets trapped in the 3:2 MMR with planet b from the outset. Because planets f \& g are inserted simultaneously and more massive planet migrate faster \eq{tm}, planet g catches up with planet f and gets trapped in the 4:3 MMR. At the end of this simulation, the period ratios of every adjacent two planets are in 3:2 except planet f \& g (4:3), fulfilling the requirement of Objective~I. 

Objective~II is to separate the b/c/d subsystem from the 3:2 and 3:2 period ratios to 8:5 and 5:3 period ratios. As a first attempt, we disperse the disc to see if we are able to evolve the planet system naturally. During disc dispersal, the semi-major axis damping and eccentricity-damping forces decrease along with the profile of the disc exponentially on a time-scale $\tau_{\mathrm{d}}=10^5\,\mathrm{yr}$. After $20\tau_{\mathrm{d}}$ we run the post-disc simulation (Stage~III) where only the stellar tidal force operates on the planets.

\Fg{model2_separate} shows the post-disc simulation. 
The eccentricities of the planets decrease with time because of tidal dissipation during the simulation, while the 3BR angles keep librating around values near $\pi$. The libration centres of the 3BR phase angles of the last adjacent planet triplets lie around $150^{\circ}$ due to the perturbation from non-adjacent resonances \citep{SiegelFabrycky2021}. Due to tidal dissipation, the gravitational potential energy of this system is decreasing while angular momentum is conserved. Therefore, angular momentum transfers outwards, from inner planets to outer planets and the period ratios of every adjacent planet pair increase together \citep{BatyginMorbidelli2013,PapaloizouEtal2018,GoldbergBatygin2021}. 

However, the simulation does not result in the observed TRAPPIST-1 planetary orbital configuration.
The 3BRs that connect the planets are inconsistent with the observation, i.e., the angle $\tbr{c,d,e}{2,5,3}$ (see the notation in \se{model}) does not librate in the observation. Stellar tides increase the period ratios especially those of the outer planet pairs. Additionally, this simulation outcome results in eccentricities of planet b \& c too high compared to the other planets, which is inconsistent with TTV analysis \citep{AgolEtal2021}. 

To sum up, \modelii model succeeds in Objective~I but fails in Objective~II. Rather than separating the entire planet system, only the separation of the b/c/d subsystem after Stage~I is required. Specifically, this requires that  $\tbr{c,d,e}{2,5,3}$ needs to be broken before the disc disappears because stellar tidal torques alone cannot. \changed{Since disc migration alone rarely reproduces the resonant chain in the TRAPPIST-1 system \citep{TeyssandierEtal2021}, there must be special events that drive the expansion of the b/c/d subsystem before disc dispersal. }

\subsection{Planet d at the cavity edge (\modeliii)}
\label{sec:model3}
\cite{OrmelEtal2017} discussed one scenario to decouple the innermost two planets with the disc as the disc truncation radius moves outward due to magnetospheric rebound \citep{LiuEtal2017}, allowing the 3:2 MMR of the inner two planet pairs to break. Here, we aim to obtain the same result by migrating planet b \& c across the cavity radius during disc dispersal, \changed{motivated by that planet b and c are three times more massive than planet d (\tb{data_agol}), therefore are easier to open a gap in the disc}. Then, stellar tides might increase the period ratios of c/b and d/c \cite{PapaloizouEtal2018}. Therefore, $r_{\mathrm{c}}$ is set to be $0.023\,\mathrm{au}$, coinciding with the observed location of planet d. 

\begin{figure}
    \centering
    \includegraphics[width=\columnwidth]{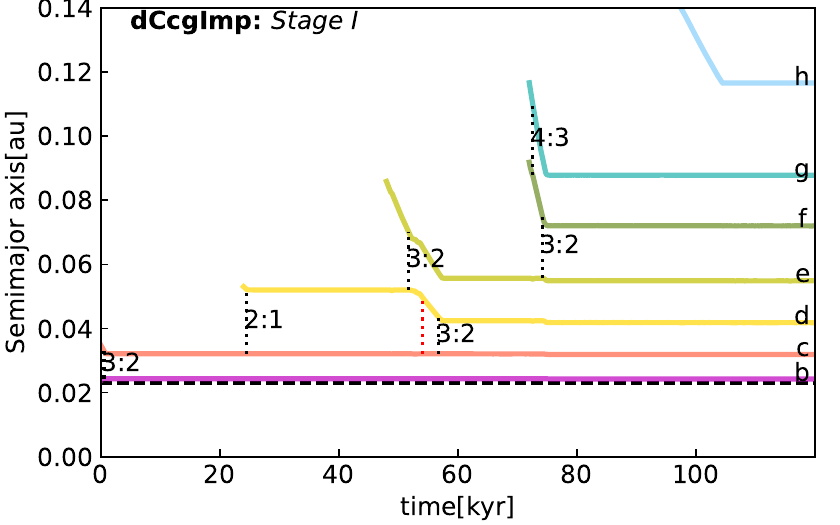}
    \caption{Similar to \fg{model2_sma}, but with $r_{\mathrm{c}}=0.023\,\mathrm{au}$. We take $\tau_{\mathrm{a,\earth}}=1.4\times10^4\,\mathrm{yr}$, $C_e=0.1$ and $A_e=1$.}
    \label{fig:model3_sma}
\end{figure}
\begin{figure*}
    \centering
    \parbox[b][][t]{0.8\columnwidth}{
    \includegraphics[width=0.8\columnwidth]{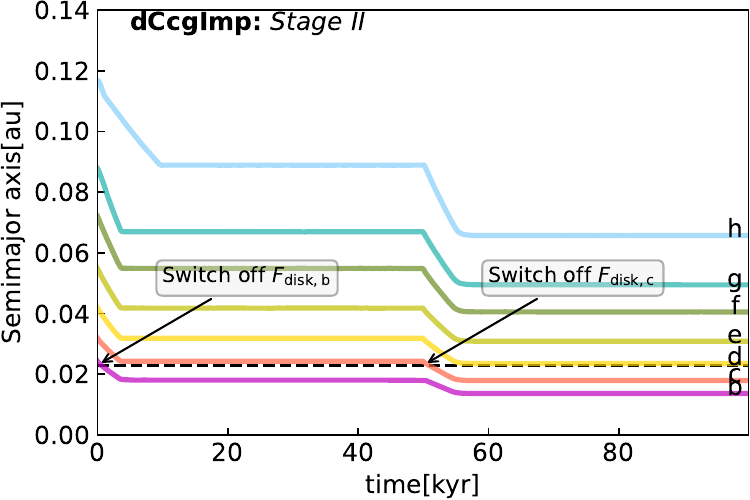}
    \includegraphics[width=0.8\columnwidth]{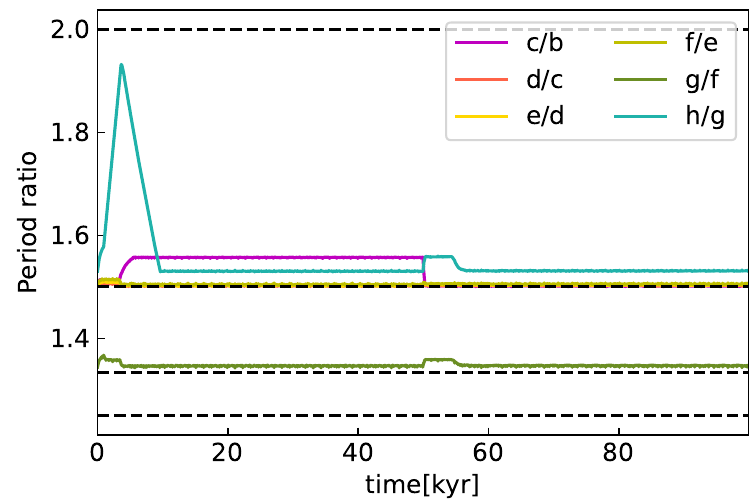}
    }
    \includegraphics[width=0.8\columnwidth]{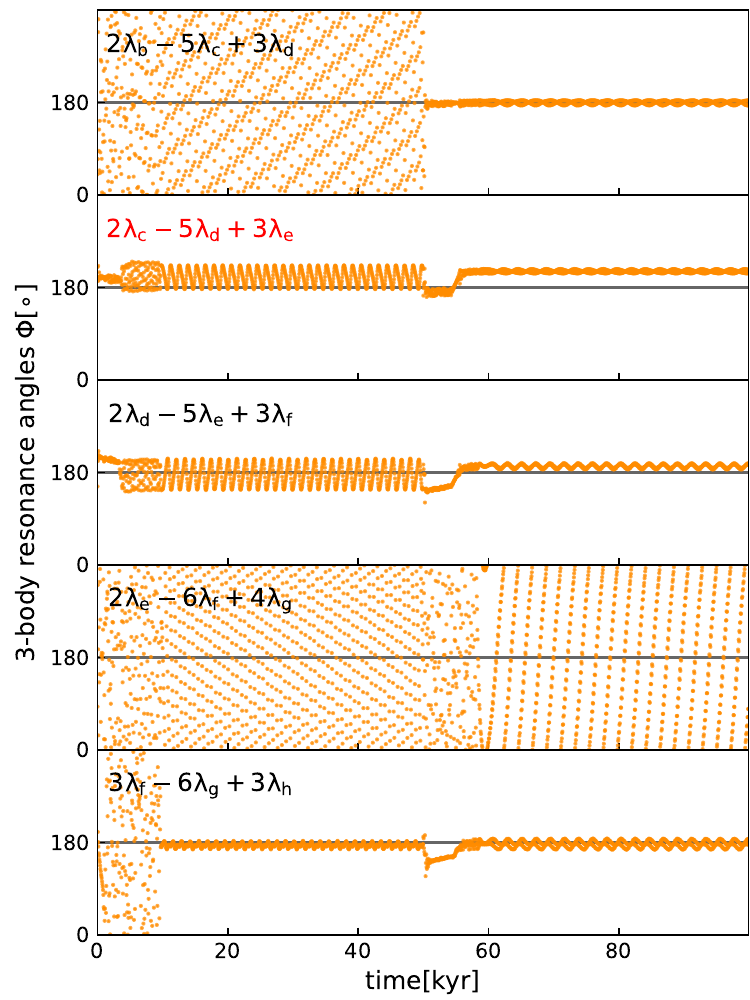}
    \caption{Stage~II of model \modeliii. In the semi-major axis plot, arrows refer to the time that we switch off the disc force on the corresponding planet. In the 3BR angle plots, we label the expression of each angle in the upper left of each panel. The label in red means that the 3-body angle does not presently librate. }
    \label{fig:model3_cross}
\end{figure*}
%In other words, disconnecting the link of incorrect 3BRs between planet c and the outer planets before 

One of our typical results for Stage~I is shown in \fg{model3_sma}. Similar to \modelii model in \fg{model2_sma}, all seven planets park themselves at the desired first order MMRs. 
\changed{To obtain the configuration that planets b and c are in the cavity while the outer planets are in the disc, we switch off the disc torque term (in \eq{eom}) on the planet at the disc inner edge, ignoring for simplicity the details of how the disc profile evolves (Stage~II).} \Fg{model3_cross} shows the simulation results for Stage~II. At the beginning of this simulation, all planets are beyond the truncation radius $r_{\mathrm{c}}$. After first switching off the disc torque on planet b, the outer planets push planet b further into the cavity until planet c has reached the cavity radius. Because its mass is small, planet h migrates slowest and is left behind the other planets. Therefore, the period ratio of planet h/g increases until the migration of planet g is stopped by the resonant interaction with the other planets in the disc. Then planet h catches up with its inner planets and the 3:2 MMR reforms. Meanwhile, stellar tides separate planet b in the cavity from exact 3:2 resonance and the period ratio of planet c/b increases to ${\approx}1.56$. Next, we switch off the disc torque on planet c, after which, the period ratio of planet h/g does not increase so much and the 3-body resonance angle $\tbr{f,g,h}{3,6,3}=3\lambda_{\mathrm{f}}-6\lambda_{\mathrm{g}}+3\lambda_{\mathrm{h}}$ librates. 

However, after planet d stops migrating at the truncation radius, stellar tides are unable to sufficiently expand the b/c/d subsystem, i.e. to increase the period ratios of planet c/b and d/c to their observed values. As can be seen from \fg{model3_cross}, the 3BR angle $\tbr{c,d,e}{2,5,3}=2\lambda_{\mathrm{c}}-5\lambda_{\mathrm{d}}+3\lambda_{\mathrm{e}}$ still librates, preventing planet c from moving further inward. Therefore, we need an additional mechanisms to break the 3BRs, specifically $\Phi_{\mathrm{c,d,e}}$, such that the planet b/c/d subsystem is allowed to expand until the observed period ratios (Objective~II). This breaking of the 3-body resonance must occur before the disc disappears as otherwise {\textit{all}} planets would undergo separation by stellar tides, similar to what is going on in \fg{model2_separate}. Therefore, a more sophisticated model is required to expand the b/c/d subsystem.

\subsection{Parameter variation for Objective~I (\modeliii)}
\label{sec:ObjectiveI}
\changed{
To assess the likelihood of fulfilling Objective~I, we conduct a sensitivity study of the model parameters during Stage I of the \modeliii model, assuming planets all start migrating outside the 2:1 MMRs of their inner planets, except for planet g, which starts just inside 3:2. The goal is to investigate whether (and which) additional planets feature such formation imprints, i.e., whether it is necessary to initialize planets inside 3:2 MMR or 2:1 MMR.
}
%We conduct a sensitivity study of the model parameters in fulfilling Objective I during Stage I of \modeliv model. 

We run a set of Stage~I simulations, varying three parameters: the migration time-scale $\tau_{\mathrm{a,\earth}}$, the eccentricity-damping prefactor $C_e$ and the eccentricity-damping enhancement at the truncation radius $A_e$. The parameter $C_e$ is taken in the range of [0.1, 1], $\tau_{\mathrm{a,\earth}}$ in the range of [5, 20] kyr and $A_e$ within [1, 40]. We always initialize planet g inside the 3:2 MMR of planet f and other planets wide of the 2:1 MMR of their inner planets. 

\changed{The simulation results are shown in \fg{model4_sensitivity}. We divide the simulation outcomes into four groups according to: 
\begin{enumerate}
    \item the final orbital configuration of planet b \& c: circles indicate that planet b \& c are connected by 3:2 MMR and triangles indicate that planet b \& c are connected by 2:1 MMR. 
    \item the final configuration of the outer planets other than planet b \& c: green symbols indicate that all outer planet pairs are consistent with Objective~I and blue symbols that at least one planet pair is connected by a wider MMR than envisioned by Objective~I (and no pair in closer MMRs). 
    \item red crosses indicate that there is at least one planet pair connected by a closer MMR than stipulated in Objective~I.
\end{enumerate}
}

In panel (a), more green and blue triangles are present at higher $\tau_{a,\earth}$ since planets are easier trapped into wider MMRs -- i.e., the 2:1 -- under conditions of slower migration. As $C_e$ increases, the green and blue triangles move to the left. Higher $C_e$ results in higher planet eccentricities, decreasing the critical time-scale to cross the resonance (consistent with the analysis in \se{MMR_ecc}). Other panels show similar patterns but the higher the $A_e$, the easier for planet c to cross the 2:1 MMR with planet b. Therefore, more circles appear at high $\tau_{\mathrm{a,\earth}}$ as $A_e$ increases. The analytical explanation can be found in \se{MMR_ecc}. All green circles satisfy the requirement of Objective~I. 
For runs labelled by blue markers, it is mostly planet h that fails to cross the 2:1 MMR. \changed{Therefore, we conclude that a formation imprint may be necessary for planets c, g and h for Objective I to be fulfilled -- its outcome is surely not universal but does also not require extremely tuned conditions. }
%In other cases where more planets are in wider MMRs, formation imprint should be made use of more cautiously. \cim{I think you need to say more..}

\begin{figure*}
    \centering
    \includegraphics[width=1.7\columnwidth]{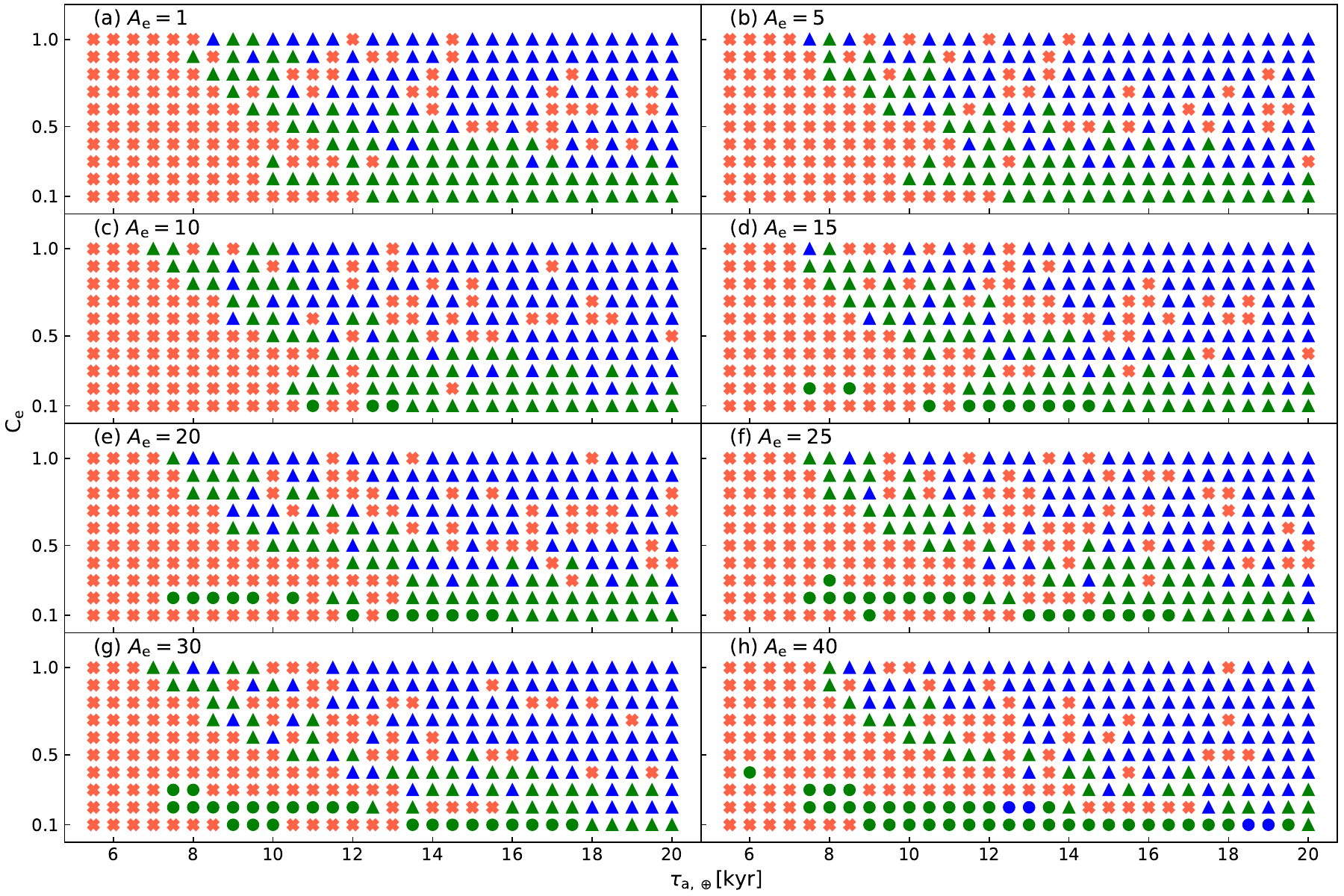}
    \caption{Sensitivity study of the \modeliii model. In all five panels the $x$-axis denotes the migration time-scale $\tau_{\mathrm{a,\earth}}$ and the $y$-axis the eccentricity-damping $C_e$, but in each panel the eccentricity-damping enhancement at the truncation radius parameter, $A_e$, differs. We label the value of $A_e$ at the top of each panel. Different symbols represent different final orbital configuration in the simulation. Circles represent that planet b \& c are connected by 3:2 MMR and triangles represent that planet b \& c are in 2:1 MMR. Red crosses indicate there is at least one planet pair in a closer MMR than the desired configuration of Objective~I, green indicates that all outer planet pairs are consistent with Objective~I and blue indicates that at least one planet pair is in an MMR wider than the desired resonance chain.}
    \label{fig:model4_sensitivity}
\end{figure*}

\changed{
Since one of our assumptions is that the disc is viscously relaxed (as described in \se{model}), the $\alpha$-viscosity parameter can be constrained using \eq{ta_n}. 
In \fg{model4_sensitivity}, green circles are centred near $\tau_{\mathrm{a,\earth}}=12\, \mathrm{kyr}$, corresponding to a value of $\alpha\sim10^{-3}$. $\alpha\sim10^{-3}$ is a typical value seen in observational studies, e.g., on the disc gas radii \citep{TrapmanEtal2020} and stellar accretion rates \citep{HartmannEtal2016}, as well as theoretical studies, e.g. on modelling the dust size distribution \citep{BirnstielEtal2018}. A higher value $10^{-2}-10^{-3}$ is suggested in the close-in region in the disc \citep{Gammie1996,CarrEtal2004}. Note that our modelling constrains the migration time-scale parameter $\tau_{a,\earth}$; when converting to the disc proprieties (like $\alpha$) additional uncertainty arises from other model-dependent parameters, i.e., $\dot{M}_{\star}$, $\gamma_I$.
}

\section{Trapping and escape of planets in 3BRs}
\label{sec:3BR_test}
\changed{This section examines how the TRAPPIST-1 planets evolve from Objective~I to Objective~II, in which the inner three planets attain their final period ratios. We postulate that this is achieved after planets b and c enter the cavity and experience a phase of inward migration. Specifically, we envision that planet c is being pushed inwards, from an initial period ratios of 3:2 (with planet d) to the final ratio of 5:3. The rationale behind this scenario is that if the $\tbr{b,c,d}{2,5,3}$ 3BR is maintained during this process, the inner planets will naturally end up near their observed period ratios.}

\begin{figure}
    \centering
    \includegraphics[width=\columnwidth]{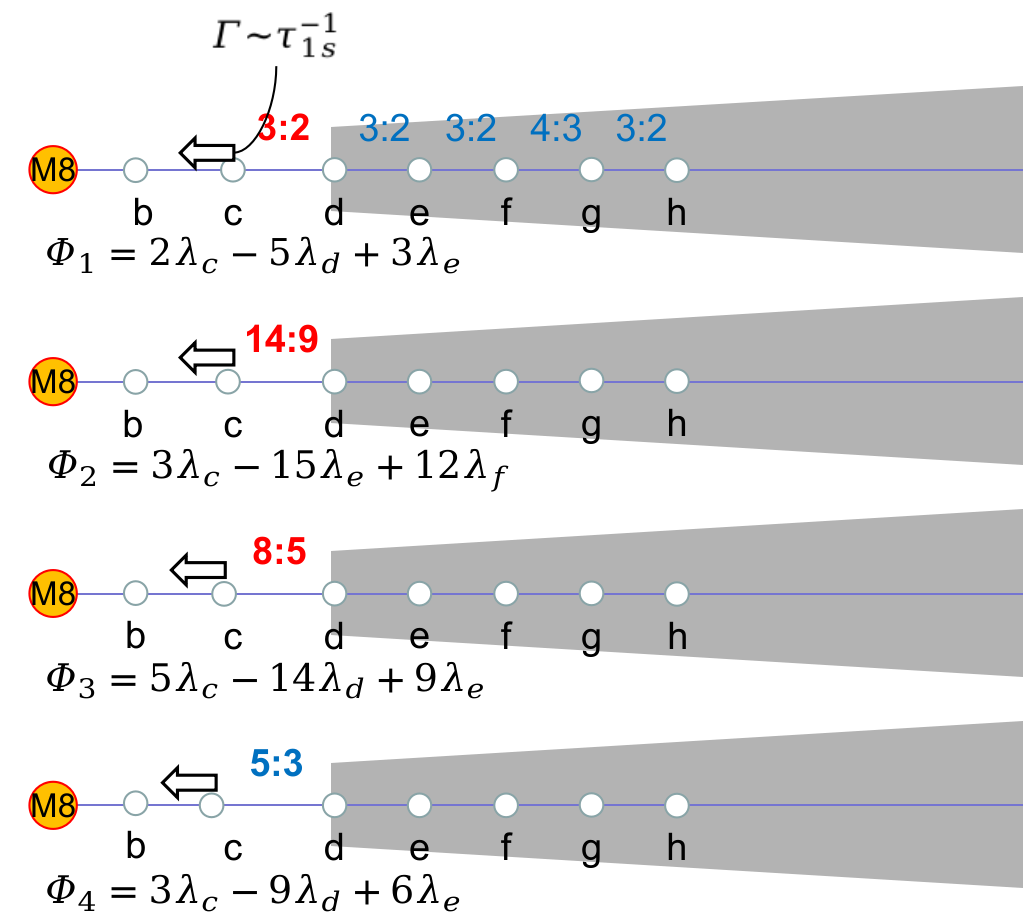}
    \caption{Four 3BR angles in which planet c could be trapped when it migrates from its 3:2 (top) period ratio to its observed value of 5:3 (bottom). 
    %\changed{Four equilibrium state when moving planet c from its 2:3 period ratio with planet d (top) to its observed value of 3:5 (bottom).} 
    The period ratios in blue are consistent with the present observation while those in red are inconsistent. Several three-body resonances need to be passed. $\Gamma_{\mathrm{1s}}$ (or its corresponding time-scale $\tau_{\mathrm{1s}}$) represents the torque from the exterior remnant disc. }
    \label{fig:param_study_model}
\end{figure}

\begin{figure*}
    \centering
    \includegraphics[width=\textwidth]{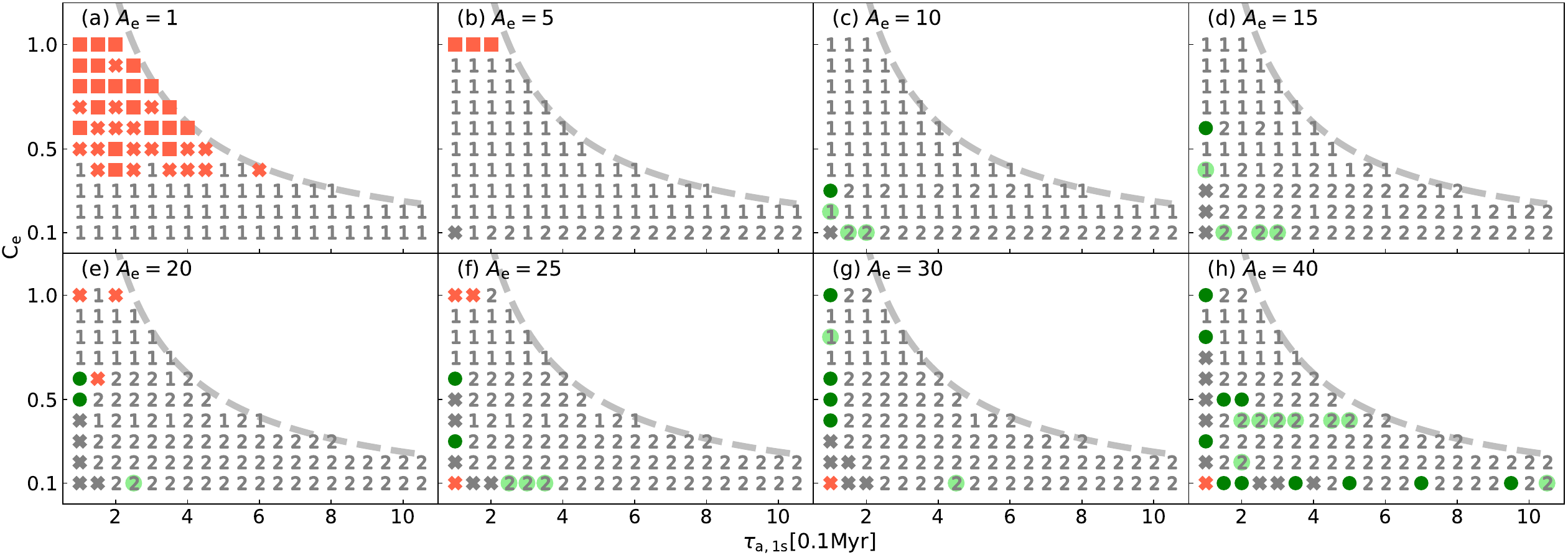}
    \caption{ \changed{Results of the Late Infall model parameter study, featuring a one-sided torque $\Gamma_\mathrm{1s}$ (here expressed in terms of a time-scale $\tau_\mathrm{1s}$) acting on planet c. Different panels refer to different values of the cavity eccentricity enhancement factor $A_e$. In each panel, the $x$-axis denotes the semi-major axis damping time-scale acting on planet c in the cavity. The $y$-axis denotes the eccentricity-damping efficiency (\eq{te_d}). The nominal migration time $\tau_{\mathrm{a,\earth}}$ is taken to be $16\,\mathrm{kyr}$. The grey dashed line is $\tau_{\mathrm{1s}}\times\,C_e=2.5\times10^5\,\mathrm{yr}$. Different markers represent different final configurations. The markers '1', '2', '3' and the green dot represent respectively the $\Phi_1$, $\Phi_2$, $\Phi_3$ and $\Phi_4$ end stages, as shown in the four panels in \fg{param_study_model}. Furthermore, grey crosses mean that planet c crosses $\Phi_4$, red crosses that at least one of outer planet pair breaks the desired MMRs, i.e., Objective~I is not yet achieved, and red squares mean that planet d crosses the disc inner edge. Finally, light-green dots indicate at least one configuration ends up in $\Phi_4$ in simulations with larger $\tau_{\mathrm{a,\earth}}$. } }
    \label{fig:param}
\end{figure*}

However, while (gently) pushing planet c inwards, it needs to overcome other 3BR with the outer planets. We list four 3BRs that may halt the inward migration of planet c and therefore the expansion of the b/c/d subsystem. In order of increasing period ratio $P_{\mathrm{d}}/P_{\mathrm{c}}$, \changed{as illustrated in \fg{param_study_model}:}
\begin{itemize}
    \item $\Phi_1\equiv\tbr{c,d,e}{2,5,3}$ with period ratio $P_{\mathrm{d}}/P_{\mathrm{c}}=1.5$;
    \item $\Phi_2\equiv\tbr{c,e,f}{3,15,12}$ (non-adjacent) with period ratio $P_{\mathrm{d}}/P_{\mathrm{c}}=1.556$; 
    \item $\Phi_3\equiv\tbr{c,d,e}{5,14,9}$ with period ratio $P_{\mathrm{d}}/P_{\mathrm{c}}=1.6$; 
    \item $\Phi_4\equiv\tbr{c,d,e}{3,9,6}$ with period ratio $P_{\mathrm{d}}/P_{\mathrm{c}}=1.667$.
\end{itemize}
The last 3BR is the observed period ratio while the first three need to be crossed. % before the proto-planet disc is gone. 
We therefore search for parameter combinations that can break the first three 3BRs but not $\Phi_4$. As a comprehensive (analytical) investigation of 3BR is beyond the scope of this work (but see some comments in \se{analysis}), we instead solve the problem numerically in \se{3b-trapping}.% to obtain a preliminary understanding on 3BR trapping. 
That is, we vary the torque on planet c, as well as the disc parameters, to assess the conditions for which planet c crosses $\Phi_1$, $\Phi_2$ and $\Phi_3$ but stays in $\Phi_4$.
Based on these findings, in \se{model4}, we implement a complete scenario for the emergence of the TRAPPIST-1 dynamical configuration. In \se{oneside}, we comment on the nature of the inward torque that planet c experiences.

\subsection{Crossing and trapping of the three-body resonances (\modeliv)}
\label{sec:3b-trapping}
% \corem{For each parameter combination of $A\mathrm{e}$, $C_e$ and $\tau_{a,\earth}$, we can access \fg{model4_sensitivity} to consult which planets need to employ formation imprint in order to explain the initial condition (Objective~I) in this section. We vary different parameters and classify the final state into different groups.}\ccc{do you need these sentences?}

\changed{We conduct a parameter study to assess the conditions necessary for the above described scenario, varying the torque on planet c $\tau_\mathrm{1s}^{-1}$, the nominal migration time $\tau_{a,\oplus}$ (see \eq{tm}), the eccentricity-to-semi-major axis damping parameter $C_e$ (see \eq{te_d}), and the eccentricity-damping boost at the disc inner edge ($A_e$; see \eq{fe}). The numerical simulations are conducted with all seven planets: five in the disc (d, e, f, g, and h) and two (b and c) in the cavity. All planets are initialized with a configuration that is consistent with Objective~I. Next, we employ a negative torque on planet c in the cavity to break the 3BRs. We record the outcome of the simulation (see below). We conduct a parameter study varying $\tau_{\mathrm{1s}}$ in the range $[10^{5},10^{6}]\mathrm{yr}$, $\tau_{\mathrm{a,\earth}}$ in the $[12, 36]\,\mathrm{kyr}$ range,  $1\le A_e \le 40$ and $0.1\le C_e \le 1$. Each simulation is integrated for ${\approx}\tau_\mathrm{1s}/3$, such that it is long enough for $P_\mathrm{d}/P_\mathrm{c}$ to increase from 1.5 to 1.667 if planet c is not trapped.}
%. the enhancement of eccentricity-damping at the cavity edge, 
% the eccentricity-to-semi-major axis damping parameter 

Escape from 3BR is qualitatively different than escape from two-body MMR. In the two-body case, under the divergent situation described above, planet c will always move away from planet d; we have verified that the critical torque is smaller than $10^{-7}\,\mathrm{yr^{-1}}$. However, the presence of a third resonant body (here planet e or f) qualitatively changes the picture. Even though the torque acting on planet c is divergent, it does not necessarily break the resonance. For example, planet c, d, and e are connected by the 3BR $\Phi_1$, corresponding to the  3:2 resonance locations of both planet c \& d and d \& e. Therefore, an increase in the period ratio $P_\mathrm{d}/P_\mathrm{c}$, due to a negative (divergent) torque acting on planet c, will likewise result in an increase of $P_\mathrm{e}/P_\mathrm{d}$, if $\Phi_1$ does not break. In the scenario that planet d and e are in the disc, Type~I (convergent) migration, however, prevents the $P_\mathrm{e}/P_\mathrm{d}$ expansion. Therefore, $P_\mathrm{d}/P_\mathrm{c}$ tends to stay near 1.5.

\changed{The results of our parameter study are displayed in \fg{param}. We present the final configurations from the simulations in which $\tau_{\mathrm{a,\earth}}=16\,\mathrm{kyr}$, and vary $A_e$ (panels), $\tau_{\mathrm{1s}}$ ($x$-axis) and $C_e$ ($y$-axis). Empirically, as $C_e$ increases from 0.1 to 1.0, $\Phi_2$ becomes stronger and is more capable to trap planet c in the cavity (MMRs hold similar characteristics, which is analyzed in \se{MMR_ecc}). It is found that planet c will always strand in either $\Phi_1$ or $\Phi_2$ when the combination $\tau_\mathrm{1s}\times\,C_e=\tau_\mathrm{c}$ exceeds a certain value. Therefore, we limit $\tau_\mathrm{c}=2.5\times10^5\,\mathrm{yr}$, which is indicated by the grey dashed line in \fg{param}. }

\changed{Different markers represent different final configurations in the simulations. At the end of each simulation, we first check whether the outer planets are still in the desired MMRs; if not we mark them by red crosses. Second, we check whether planet d enters the disc cavity; if so we mark them by red squares. In both scenarios, Objective~I has failed to materialize. Most red crosses and red squares are in the panel where $A_e=1$ and $C_e$ is relatively high, because the eccentricity of planet d is not sufficiently damped. Then, planet d will cross the disc inner edge when its eccentricity becomes as large as the migration barrier width ($e\ga\,\Delta$), or planet e crosses the 3:2 MMR due to perturbation from eccentric planet d. If planet c is trapped by $\Phi_1$, $\Phi_2$, $\Phi_3$ or $\Phi_4$, we mark the corresponding simulations with '1', '2', '3' or a green dot respectively. Otherwise, if planet c crosses all four 3BR listed here, we mark the outcome by a grey cross. }

\begin{figure*}
    \centering
    \includegraphics[width=0.66\columnwidth]{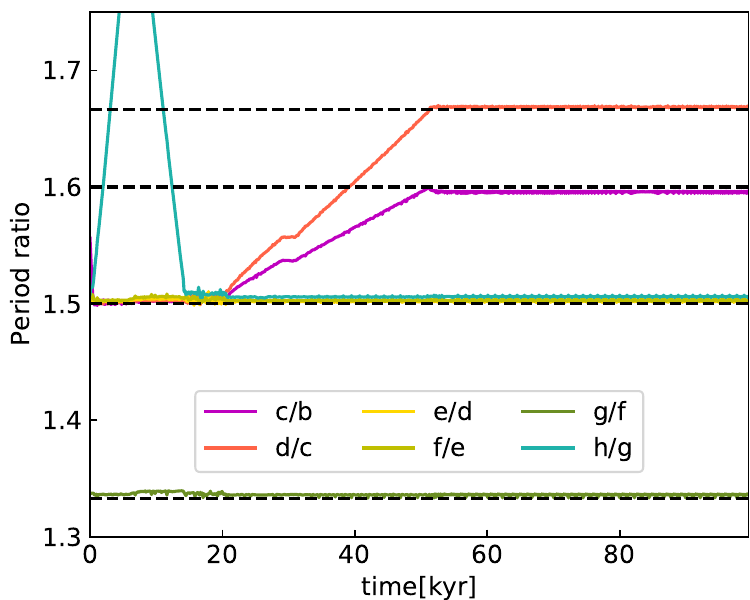}
    \includegraphics[width=0.66\columnwidth]{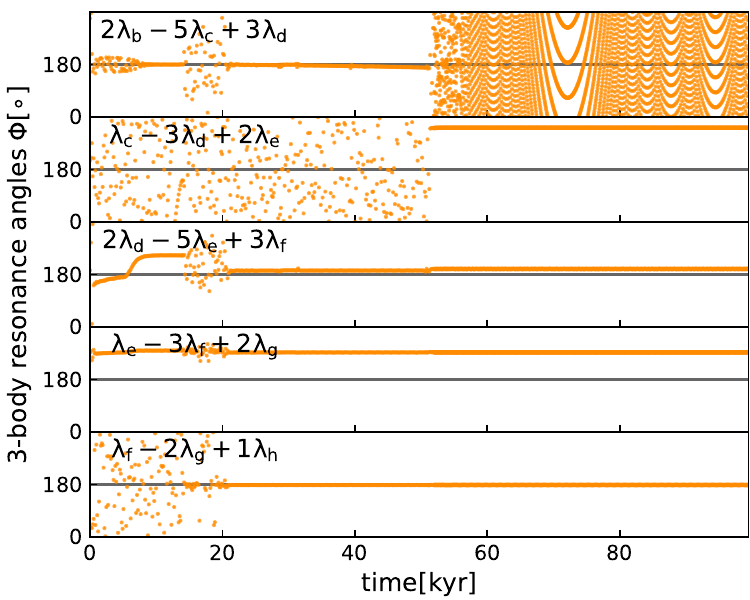}
    \includegraphics[width=0.66\columnwidth]{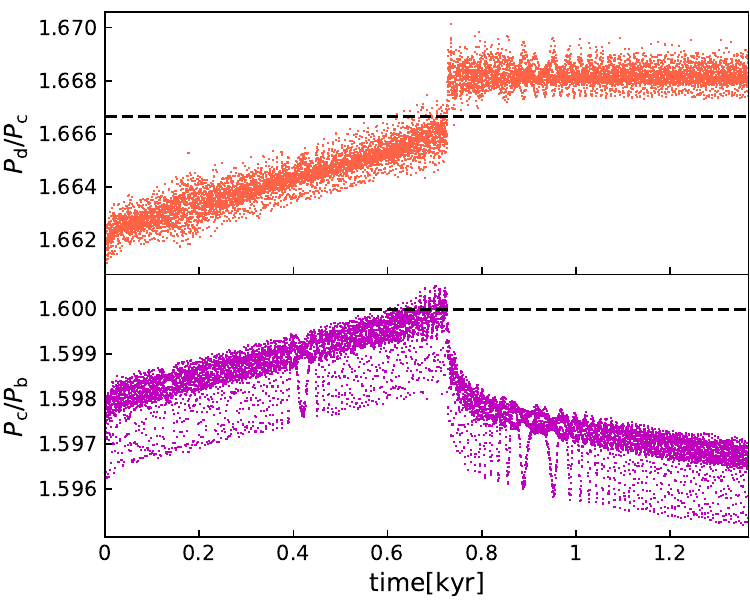}
    
    \caption{ \changed{ The period ratio and 3BR angles evolution in one of the systems drawn from \fg{param}, where $A_e=15$, $C_e=0.6$ and $\tau_\mathrm{1s}=0.1\, \mathrm{Myr}$. In the left panel, the period ratios 1.66, 1.60, 1.5, 1.33 are indicated by the four dashed horizontal lines from the upper to the bottom. In the middle panel, we label the expression of each 3BR angle. In the right panel, we zoom in on the period ratios $P_\mathrm{c}/P_\mathrm{b}$ and $P_\mathrm{d}/P_\mathrm{c}$ around the time where the b/c/d 3BR breaks. }}
    \label{fig:picked}
\end{figure*}

\changed{\Fg{param} provides some insights into the physical properties of 3BRs. In each panel, planet c tends to be trapped in $\Phi_1$ (marker '1') at high $C_e$. As $C_e$ decrease down to 0.1, $\Phi_2$ (marker '2') replaces it in most simulations. It indicates that $\Phi_1$ becomes weaker if more efficient eccentricity damping is adopted. When increasing the one-side torque on planet c in the cavity, planet c breaks through $\Phi_2$ and gets trapped in $\Phi_4$ (green dots) in several simulations. These simulations reveal two points. First, the boundary between trapping in $\Phi_2$ and $\Phi_4$ is not so clear, hinting that the trapping process of planet c is stochastic. Second, planet c is either trapped by $\Phi_4$ or just crosses $\Phi_4$ after $\Phi_2$, rather than be trapped by $\Phi_3$. It implies that $\Phi_3$ is much weaker than $\Phi_2$ and $\Phi_4$. Furthermore, $\Phi_2$ and $\Phi_4$ may have comparable strengths to trap planet c . Otherwise, more green dots would have popped up at the locations now occupied by grey crosses. }

\changed{We give one possible reason for the stochastic nature of 3BR trapping/escaping. Apart from $\Phi_1$, $\Phi_2$, $\Phi_3$ and $\Phi_4$, there are $9\times4$ other 3BRs involving the outer five planets that could trap planet c at the four period ratios $1.5$, $1.556$, $1.6$, $1.667$. For instance, $\tbr{c,d,g}{3,6,3}$ is able to trap planet c at $P_\mathrm{d}/P_\mathrm{c}=1.667$. Some of these 3BRs feature several different libration centers when differing the initial conditions slightly \citep{TamayoEtal2017}. All these different 3BRs or the same 3BR with different libration centers may have distinct strengths to trap planet c in the cavity. If the initial condition is different, planets may end up in a different configuration, thus contributing to the stochastic behavior. }

\changed{At the beginning of all of these simulations, $P_\mathrm{c}/P_\mathrm{b}$ increases in sync with $P_\mathrm{d}/P_\mathrm{c}$ because 3BR 'bind' planet b/c/d together. However, in most of the simulations where planet c is trapped by $\Phi_4$ as a final state, $P_\mathrm{c}/P_\mathrm{b}$ is slightly smaller than 8/5. We present an example in \fg{picked}. It shows that when $P_\mathrm{d}/P_\mathrm{c}$ is about to arrive at 5/3, planet c undergoes an inward resonant repulsion \citep{LithwickWu2012} which breaks the b/c/d resonance. Thereafter, $P_\mathrm{c}/P_\mathrm{b}$ decreases to $1.595$ as shown on the right panel in \fg{picked}. This intermediate configuration is a natural output from our model, which is regarded as an initial configuration by \cite{PapaloizouEtal2018}. Long term tidal dissipation then expand $P_\mathrm{c}/P_\mathrm{b}$ back to ${\approx}8/5$ \citep[\se{model4}, cf.][]{PapaloizouEtal2018}, which is consistent with the current observation. }

\changed{Disc evolution may also help to move the TRAPPIST-1 planets into their observed configuration. We run five sets of simulations similar to \fg{param}, but with $\tau_{\mathrm{a,\earth}}$ equal to $20\,\mathrm{kyr}$, $24\,\mathrm{kyr}$, $28\,\mathrm{kyr}$, $32\,\mathrm{kyr}$ and $36\,\mathrm{kyr}$ (not shown). In several of these simulations with higher $\tau_{\mathrm{a,\earth}}$ planet c becomes trapped by $\Phi_4$ and we mark them by light-green dots in \fg{param} together. 
Therefore, if disc evolution operates during the migration planet c may avoid trapping in $\Phi_1$ and $\Phi_2$ to end up in $\Phi_4$.}

%Therefore, when the disc disperses ($\tau_{\mathrm{a,\earth}}$ increases), the cases where planet c is trapped by $\Phi_1$ and $\Phi_2$ in \fg{param} may yet allow become unstable and the 3BRs could still end up in $\Phi_4$.\ccc{I feel we put this a bit too strong.} \corem{In other words, when considering disc evolution, the probability of planet c being trapped by $\Phi_4$ might become higher than what \fg{param} presents.}
%\coadd{However, the hypothesis that $\Phi_i$ may yet break upon a reducing one-sided torque seems counter intuitive; to properly assess this scenario a time-dependent $\tau_{1s}$ is required.}}

\changed{In this model, we have successfully reproduced the orbital configuration of TRAPPIST-1 planets as well as their resonant relationship. However, most of the green dots feature low $C_e$ ($<0.7$) and extremely high $A_e$ (${\gtrsim}10$). Besides, the one-sided torque on planet c in the cavity also needs to be tuned. Considering all these caveats, the probability of TRAPPIST-1 analogs produced by this scenario is not very compelling. }

\begin{figure*}
    \centering
    \includegraphics[width=1.6\columnwidth]{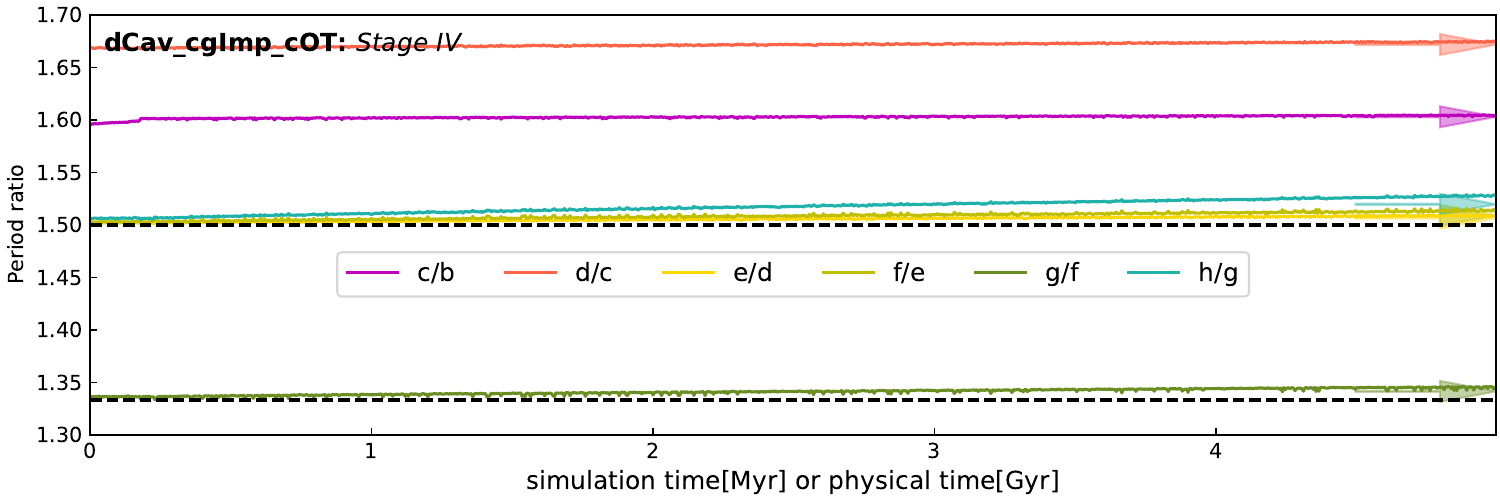}
    \includegraphics[width=0.8\columnwidth]{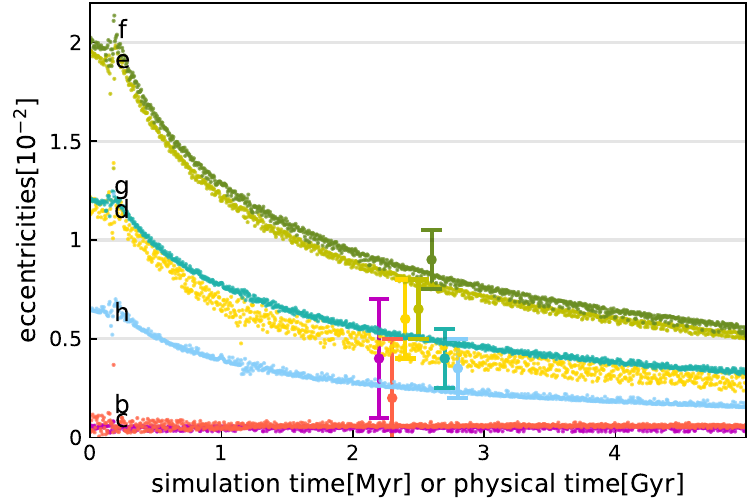}
    \includegraphics[width=0.8\columnwidth]{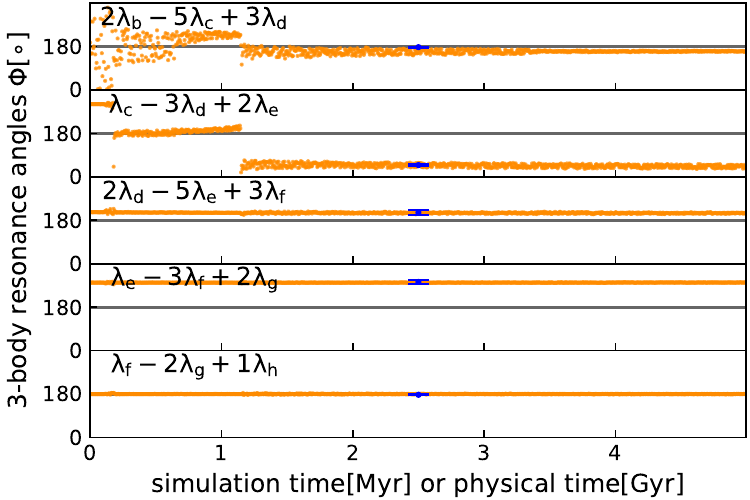}
    \caption{\changed{Long-term tidal damping simulations of the post-disc phase, illustrating how, after the completion of Objective II, the system evolves towards its present configuration. We present period ratios (top panel), eccentricities (bottom-left) and 3BR angles (bottom-right) during tidal dissipation (with $Q_{\mathrm{sim}}=0.1$), after dispersing the gas disc of the system in \fg{picked}. The figure style is the same as \fg{model2_separate}. We label the observed range of 3BR angles \citep{LugerEtal2017} on the lower right panel with blue error bars. Tidal damping reforms the $\tbr{b,c,d}{2,5,3}$ 3BR and damps the eccentricities towards the observationally-inferred values.}}
    \label{fig:model4_tide}
\end{figure*}

\subsection{Long term tidal dissipation (\modeliv)}
\label{sec:model4}
\changed{The breaking of 3BR of planets b, c, and d is a natural outcome in \modeliv model, as described in \fg{picked}. The 3BR of the innermost three planets is believed to librate \citep{LugerEtal2017, AgolEtal2021, TeyssandierEtal2021}. To reform this 3BR, long-term tidal dissipation is needed. }

\changed{We therefore run a long-term simulation, similar to \fg{model2_separate} for one particular parameter set (the same as the one in \fg{picked}). We first disperse its disc on a e-folding time-scale of $10^5\mathrm{yr}$ and run the simulation for $20\tau_\mathrm{d}$. Using the final snapshot from the disc dispersing simulation as an initial condition, we start long term tidal dissipation process.
We take $Q_{\mathrm{sim}}= 0.1$ for all planets to accelerate the simulation by $10^3$ times \citep{PapaloizouEtal2018}.}

\changed{Since the disc dispersal does not modify the orbital properties of the planets in this system, we only present the evolution of the planets after removing the disc in \fg{model4_tide}. }
The period ratio $P_{\mathrm{c}}/P_{\mathrm{b}}$ increases with time due to tidal damping during the first $\approx0.2\, \mathrm{Myr}$ simulation time ($\approx0.2\,\mathrm{Gyr}$ physical time). After that, the period ratio $P_{\mathrm{c}}/P_{\mathrm{b}}$ reaches 1.6 and $\tbr{b,c,d}{2,5,3}$ reforms. Then, the period ratio increase of the inner planets slows down because all planets are linked by a chain of 3BRs. \changed{At $t\approx2.5\,\mathrm{Myr}$ simulation time ($t\approx2.5\,\mathrm{Gyr}$ physical time)}, the eccentricities and period ratios of all planets are consistent with the inferred values by \cite{AgolEtal2021} indicated by the $1\,\sigma$ error bars in \fg{model4_tide}. The libration centre of all five adjacent 3BRs lie within the observationally-inferred values as well. 

\changed{Our simulation results can constrain the tidal quality factor of TRAPPIST-1 b.} The estimated age of TRAPPIST-1 is $7.6\, \mathrm{Gyr}$ \citep{BurgasserMamajek2017}, suggesting that the simulation in \fg{model4_tide} evolves \changed{${\approx}3$} times faster than the real TRAPPIST-1 system. We also run Stage~III simulations that apply tidal torques on (i) only planet b; (ii) only planet b and c. The results are consistent with the simulation that applied the tidal torque on all seven planets. The tidally induced expansion is therefore dominated by the tidal dissipation of planet b \citep[cf.][]{PapaloizouEtal2018}. Since $Q_\mathrm{sim}=3Q/(2k_2)$, the chosen simulation constrains the tidal quality factor of planet b as \changed{$Q_\mathrm{b}\approx200 k_2$}. This value falls within the upper limit of $5,000k_2$ calculated by \citet{BrasserEtal2019}. \changed{Note that the bulk planetary eccentricities just after removing the surrounding disc are proportional to $\sqrt{C_e}$ \citep{GoldreichSchlichting2014,TeyssandierTerquem2014,TerquemPapaloizou2019}. The time one system spends in the tidal dissipation simulation to fit the observed TRAPPIST-1 system is positively correlated with the initial bulk eccentricities. Our model features that $C_e\lesssim0.6$, and therefore the corresponding constraint on the tidal quality factor is $Q_\mathrm{b}\gtrsim\,200 k_2$.}

\changed{
In \fg{model4_tide}, long term stellar tidal damping does not alter the period ratios very much compared to \fg{model2_separate}. 
Since the initial eccentricities of planets b and c are ten times smaller than in \fg{model2_separate}, stellar tidal forces on planets b and c are also smaller (\eq{tidal_force}) hence dissipating the energy of the system on a longer time-scale. Therefore, the departures from exact commensurability increase more slowly in \fg{model4_tide}. }

\section{Early cavity infall (\modelv)}
\label{sec:new_model}
% \changed{The former part of this paper features that planet b and c enter the cavity after the planets' sequential migration into a first-order resonance chain. However, the order of the two process could be reversed. In this section, we assume planet b and c were in the disc and get trapped in 3:2 MMR initially. As they grow near the disc inner edge or the disc become thinner, they are able to open gaps in the disc and enter the cavity \citep{AtaieeKley2021}. Outer planets are formed afterwards as the sketch \fg{new_scenario} illustrates. We specify the one-side negative torque operating on planets in the cavity using 'Lindblad torque'. This model is named as \modelv.
% }

\changed{In the preceding, we have assumed that planets b and c crossed the cavity only after the emergence of a first order resonance chain, separating Objective I and II. Objective II is the most challenging to meet, as this requires planet c to cross the $\Phi_1$--$\Phi_3$ 3BRs but not to overshoot $\Phi_4$. However, in reality these stages may be intertwined, such that planets b and c may enter the cavity before the outer resonance chain has established, that is, before (some of the) $\Phi_i$ even exist. For this scenario, we hypothesize that planet b and c, being larger, were able to open partial gaps in the disc that caused them to be disconnected from the disc \citep{AtaieeKley2021}, whereas planet d, being smaller, did not suffer this fate and stayed at the cavity edge. While in the cavity we assume, as before, that planet c experiences an inward migration torque, which causes it to drift away slowly from planet d. }

\changed{We only concentrate on how likely it is to evolve to Objective~II from Objective~I, because our analysis in \se{ObjectiveI} (see \fg{model4_sensitivity}) is still roughly applicable. The model is illustrated in \fg{new_scenario}. At $t=0$, we assume that planets b and c and d are initially in 3:2 MMR with $d$ at the cavity edge. Subsequently, planets e, f, g, and h enter the resonant chain one after another, each separated by an interval time $\tau_\mathrm{pl}$ -- the planet formation time. Meanwhile planet c experiences a negative torque, expressed in terms of a migration time-scale $\tau_\mathrm{1s}$. We now introduce a specific model for the nature of this "one-sided" torque (\se{oneside}), linking it to the the scaling of nominal Type~I semi-major axis damping time-scale $\tau_{a,\oplus}$, i.e., $\tau_\mathrm{1s}$ is no longer a free parameter. Furthermore, we no longer vary the eccentricity-damping enhancement parameter at the cavity edge, $A_e=1$, leaving us to explore $C_e$, $A_a$, $\tau_{a,\oplus}$, and $\tau_\mathrm{pl}$ in \se{model5}.}

\begin{figure*}
    \centering
    \includegraphics[width=1.8\columnwidth]{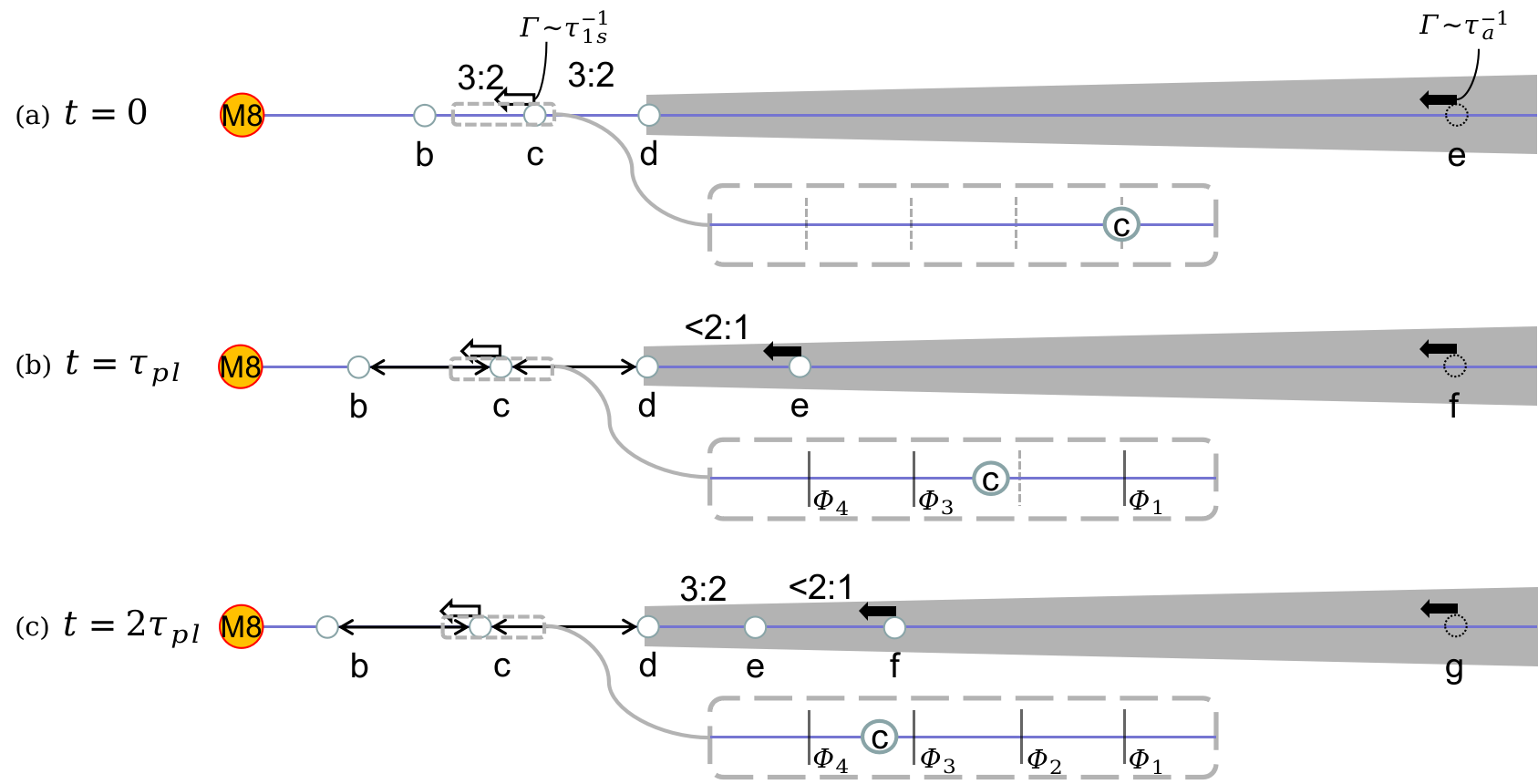}
    \caption{ \changed{Schematic of the Early Cavity Infall model (\modelv). At $t=0$ (upper panel), planets b and c have already entered the cavity and planet d has migrated to the inner disc edge. Planet b, c, and d are in 3:2 MMRs. Planet e is still forming or migrating (on time-scale $\tau_{a}$) far out in the disc. Planet c is pushed by the external disc inward on time-scale $\tau_\mathrm{1s}$. At $t=\tau_\mathrm{pl}$ (middle panel; $\tau_\mathrm{pl}$ is the planet formation interval time) planet e enters the resonant chain and is inserted inside the 2:1 MMR of planet d. At this point, planet f is forming or migrating in the outer disc. At $t=2\tau_\mathrm{pl}$ (bottom panel), planet f ends up inside the 2:1 MMR of planet e. Similarly, planet g and h are inserted after every $\tau_\mathrm{pl}$, inside 3:2 MMR and 2:1 MMR respectively. In each panel, we zoom in on the region around planet c and indicate the four 3BRs introduced in \se{3BR_test} that can trap planet c by solid vertical lines. At early times some of the $\Phi_i$ may not yet exist (dashed vertical lines; unlabelled), allowing planet c to cross these locations. 
    } }
    \label{fig:new_scenario}
\end{figure*}

\subsection{Disc repulsion on planets in the cavity}
\label{sec:oneside}
The Lindblad torque is capable to drive planets in the cavity inward.
It arises due to gravitational exchange at the  Lindblad resonance locations in the disc \citep{Ward1997}. In the case that the planet is located in the inner disc cavity, only those outermost Lindblad resonances, which still lie in the disc will contribute. Because the nominal migration rate (for an 1 Earth mass planet) and the distant migration rate both scale with disc mass, the rates are linked through: 
\begin{equation}
    \label{eq:lindblad}
    \tau_{\mathrm{1s,\earth}}^{-1}
    = \tau_{\mathrm{a,\earth}}^{-1} \sum_{m\ \textrm{in disc}} f_{m}
    = 0.0054 \tau_{a,\earth}^{-1}
\end{equation}
where $f_{m}$ is the torque, arising from the $m$th order Lindblad resonance, normalized to the nominal Type~I migration torque ($L/2\tau_{a,\earth}$, where $L$ is the angular momentum of the planet), and the last expression assumes only $m=1$ contributes (see below). The Lindblad resonance condition for an orbit of $m\mathrm{th}$ order torque reads:
\begin{equation}
    \label{eq:lindblad_loc}
    m[\Omega(r)-\Omega_{\mathrm{p}}]=\pm\kappa\sqrt{1+m^2h^2},
\end{equation}
where $\Omega(r)$ is the Keplerian velocity at distance $r$, $\Omega_{\mathrm{p}}$ is the Keplerian velocity on the planet orbit and $\kappa$ is the epicycle frequency. We follow the calculation by \cite{Ward1997} and plot the value of $f_{m}$ for $m$ from 1 to 50 in \Ap{Ward_calcu}.

In the case of the TRAPPIST-1 planets, after planets b and c have entered the cavity, planet d resides near the cavity radius location. As the period ratios of planets d and b are larger than 2:1, there is no one-sided Lindblad torque from the outer disc on planet b. On the other hand, the period ratio of planets c and d exceeds 1.5 but is less than 2. Therefore, only the outer 2:1 Lindblad resonance of planet c falls in the disc and the relevant order is $m=1$. The value of $f_{1}$ for the outer Lindblad resonance is 0.0054. For instance, for $\tau_{a} = 10\,\mathrm{kyr}$ we arrive at $\tau_\mathrm{1s}=1.85\,\mathrm{Myr}$. \changed{Although $\tau_\mathrm{1s}$ is long, the time to cross a distance from --say-- the 3:2 to the 5:3 resonance location is lower by about a factor ${\sim}10$. Therefore $\tau_\mathrm{1s}$ may compete with the planet formation interval time-scale $\tau_\mathrm{pl}$. }

\subsection{Parameter study (\modelv)}
\label{sec:model5}
% \changed{\ccc{this text reads well. But most of it is already described in the new par I wrote above. Please merge.}
% We only concentrate on how likely it is to evolve to Objective~II from Objective~I, because our analysis for \fg{model4_sensitivity} is still roughly applicable here. Different from \se{3b-trapping}, we always fix $A_e=1$ in this section. For every single simulation, we initialize planets b and c in the cavity and d on the disc inner edge. Each adjacent pair of planets are trapped by 3:2 MMR and the three are linked by 3BR at the beginning. Then the outer planets are inserted into the simulation one by one, slightly outside their desired MMRs. The time span of these planets' inserting physically means the planet formation span $\tau_\mathrm{pl}$, which is regarded as a simulation parameter. }

\begin{figure*}
    \centering
    \includegraphics[width=2\columnwidth]{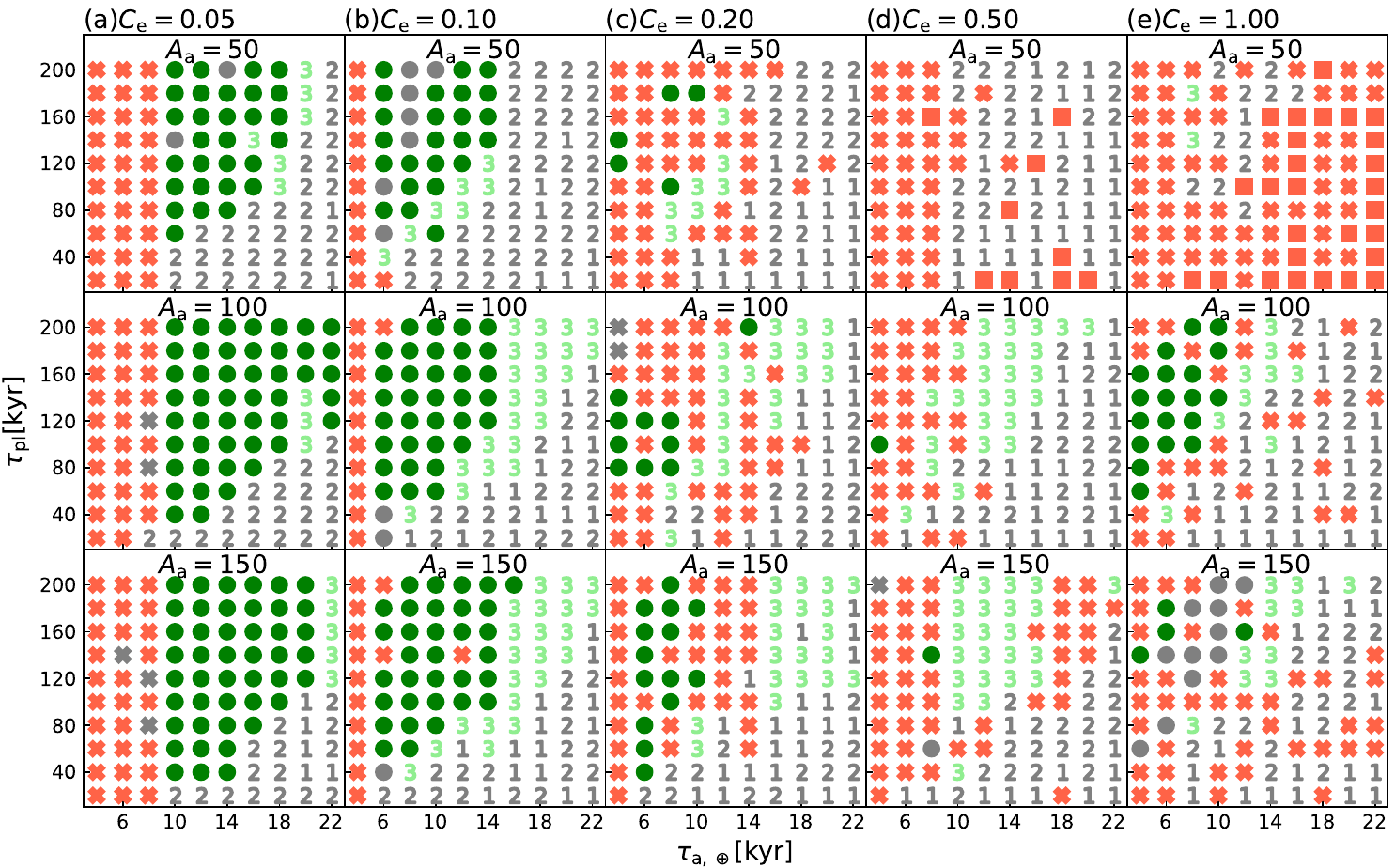}
    \caption{\changed{Final orbital configurations produced by the Early Cavity Infall model (\modelv). Different columns and different rows refer to different eccentricity-damping parameter $C_e$ and migration threshold height $A_a$. In each panel, the $x$-axis denotes the Type~I semi-major axis damping time-scale $\tau_{a,\earth}$, the $y$-axis denotes the planet formation span $\tau_\mathrm{pl}$. Different markers represent different final configurations, following the same style as \fg{param}. We additionally mark the configurations in which planet c stays in $\Phi_4$ with $P_\mathrm{c}/P_\mathrm{b}<1.53$ or $P_\mathrm{c}/P_\mathrm{b}>1.63$ by grey dots. }}
    \label{fig:model_new}
\end{figure*}

\changed{{We conduct a parameter study to assess the conditions needed to obtain the desired resonance configuration. Apart from varying the interval time $\tau_\mathrm{pl}$} from 20\, to 200\, kyr, we also vary the eccentricity-to-migration damping parameter $C_e$ from 0.05 to 1, the migration threshold barrier parameter $A_a$ from 50 to 150, and the nominal planet migration parameter $\tau_{a,\earth}$ from $4\,\mathrm{kyr}$ to $22\,\mathrm{kyr}$.  We let the system evolve until ${\sim}\tau_\mathrm{1s}$ (\eq{lindblad}). The results are shown in \fg{model_new}. We mark the final configurations following the same style as \fg{param}. We also mark the configurations where planet c stays in $\Phi_4$ but where the period ratio has deviated by some amount from 8:5 ($P_\mathrm{c}/P_\mathrm{b}<1.53$ or $P_\mathrm{c}/P_\mathrm{b}>1.63$) with grey dots. }

\changed{In this Early Cavity Infall model the new parameter $\tau_\mathrm{pl}$ sensitively determines the resonant structure. Although we assume $\tau_\mathrm{pl}$ is the same for all outer planets, what matters chiefly is the arrival time of planets e and f. If $\tau_\mathrm{pl}$ is too long, planet c would migrate so far such that $P_\mathrm{d}/P_\mathrm{c}>5/3$ before planet e enters the resonant chain. Our study does not cover such long $\tau_\mathrm{pl}$. In contrast, if $\tau_\mathrm{pl}$ is too short, planet c would not migrate very far with $P_\mathrm{d}/P_\mathrm{c}<14/9$ after the arrival of planet f, which bring about the existence of $\Phi_2$. Then it is unlikely for planet c to overcome $\Phi_2$. We are hence searching for the sweet spot where planet c finally stays at $\Phi_4$. }

\changed{Starting from the bottom-right corner of each panel in \fg{model_new}, the one-side torque is relatively weak and $\tau_\mathrm{pl}$ is short. Planet c mostly stays in $\Phi_1$ or $\Phi_2$. With increasing $\tau_\mathrm{pl}$ and decreasing $\tau_{a,\earth}$, planet c can migrate over larger distances in the cavity before the outer planets appear. Hence, planet c increasingly ends in $\Phi_3$ and $\Phi_4$. Interestingly, in the left two columns green dots are more frequent than marker '3' and there is usually a clear boundary between the two markers. The corresponding one-sided torque at the boundary therefore represents the strength of $\Phi_3$. It implies that $\Phi_4$ is sturdier than $\Phi_3$, which provides a wide range of parameter space to overcome $\Phi_3$ but stay in $\Phi_4$. For this reason, we colour '3' light green -- if (for some reason) planet c manages to escape $\Phi_3$, it is still likely to end up in $\Phi_4$. } 

% \changed{To clarify, it is not necessary to recruit planet formation imprint, which prefers short $\tau_\mathrm{pl}$ only, for planet e and f to cross barrier MMRs, as illustrated in \se{ObjectiveI}. Therefore, the initial condition used here does not conflict with the former analysis on Objective~I, .}

\changed{When $\tau_{a,\earth}$ becomes too small, some planets in the disc will break their desired MMRs. As shown in \fg{model_new} when $C_e<0.2$, there are some red crosses on the left and a clear boundary between red crosses and green dots. The corresponding disc migration torque at the boundary indicates the critical torque, above which the desired configuration of outer planets (d, e, f, g, and h) will break. As $C_e$ increases from 0.05 to 0.1, the desired configuration of outer planets becomes more stable (the boundary between red crosses and green dots moves to lower $t_a$). This trend is consistent with \fg{model4_sensitivity}, and is analysed in \se{MMR_ecc}. }

\changed{However, when increasing $C_e$ to higher values, the boundary between red crosses and green dots becomes less sharp and the appearance of red crosses becomes stochastic. The reason is that with higher $C_e$ planet eccentricities are higher, which renders phenomena like resonance crossing more stochastic \citep{OgiharaKobayashi2013}. Therefore, we see more and more red crosses distributed in the right columns of \fg{model_new} (higher $C_e$). }

\changed{We test the influence of the migration threshold barrier $A_a$ by three different values, 50, 100, 150 in \fg{model_new}. If $A_a$ is small, it shows more red squares, where planet d enters the cavity, when $C_e$ is high. When increasing $A_a$, the results of $A_a=100$ is largely similar to $A_a=150$, signifying a threshold above which planet d stays at the disc edge. }

\changed{We finally comment on the meaning of the grey dots in \fg{model_new}. As a rule of thumb we see that, similar to \fg{picked}, when planet c reaches $\Phi_4$ the 3BR $\tbr{b,c,d}{2,5,3}$ breaks, whereafter it will reform due to long term tidal dissipation (\fg{model4_tide}). However, it is found that in some of the simulations, the 3BR $\tbr{b,c,d}{2,5,3}$ already breaks early. Two physically different mechanisms result in grey dots: (a) If $\tau_\mathrm{pl}$ is so long that $\Phi_4$ is reached before planets f and g appear, the outer planets f and g will further push planet c slightly inward. This case applies to the top left two panels in \fg{model_new}, where $\tau_\mathrm{pl}$ is long and $A_a$ is small. (b) The 3BR $\tbr{b,c,d}{2,5,3}$ breaks early when $\Phi_3$ is crossed or when a new planet enter the resonant chain. Since planet c is still repulsed inward by the Lindblad torque, $P_\mathrm{c}/P_\mathrm{b}$ decreases towards ${\approx}1.5$. We have run long-term tidal dissipation simulation starting with the analog satisfying $P_\mathrm{c}/P_\mathrm{b}\approx1.53$, but found that it fails to reform the 3BR $\tbr{b,c,d}{2,5,3}$.}

\changed{When adopting low $C_e$ ($\lesssim0.20$) the low eccentricities of the outer planets render the system more stable, preventing resonance crossing. Interestingly, there is another green dots island on the panel where $C_e=1$ and $A_a=100$. We do not have a clear explanation for the occurrence of so many green dots in this panel. Possibly, the strength of $\Phi_3$ is non-linearly dependent on $C_e$, such that it is strongest at $C_e\approx0.5$ and breaks again when $C_e=1$. }

\changed{To achieve the observed TRAPPIST-1 configuration, the early cavity infall model compares favourably to the late infall model of \se{3b-trapping}. The early cavity infall model requires less parameter tuning -- we fix $A_e=1$ and link $\tau_{1s}$ to $\tau_{a,\earth}$ -- and the values for the $A_a$ and $\tau_\mathrm{1s}$ parameters are natural or in line with formation models (see \se{discuss_models}). In particular for low $C_e$, the likelihood of a successful outcome is quite high.}
%The value of the migration threshold barrier $A_a$ is consistent with the calculation by \cite{LiuEtal2017}\coadd{, and $\tau_\mathrm{pl}\sim10^5\,\mathrm{yr}$ is consistent with formation models. 
%, quite likely that the TRAPPIST-1 analog can be  is reproduced successfully within a planet formation framework

\section{Analysis}
\label{sec:analysis}
\subsection{Trapping under convergent migration in MMRs}
\label{sec:MMR_ecc}
Through numerical simulations, we find that more efficient eccentricity-damping promotes resonance crossing (see \fg{model4_sensitivity}). The trend can be understood using the pendulum model \citep{MurrayDermott1999} for internal MMRs. Assuming small deviations from the equilibrium value for the resonance angle, the libration width of a $j{:}j-1$ MMR can be written as:
\begin{equation}
    \label{eq:libration_width}
    \Delta_a=\left(\frac{16|C_{j}| e}{2n}\right)^{1/2}a,
\end{equation}
where $n$, $e$ and $a$ are the mean motion, eccentricity and semi-major axis of the inner planet, respectively. The constant $C_{j}$ arises from the resonant part of the disturbing function \citep{MurrayDermott1999}:
\begin{equation}
    \label{eq:cr}
    C_j=\frac{m'}{M_\star} n \alpha_{j} f_{j}
\end{equation}
which is proportional to the inner planet mass $m'$. In \eq{cr}, $\alpha_{j}$ and $f_{j}$ are the inner-to-outer planet semi-major axis ratio and the direct term in the expansion of the disturbing function. The value of the most commonly used $f_{j}$ can be seen on page 334 in \citep{MurrayDermott1999}. The libration time-scale is:
\begin{equation}
    \label{eq:libration_time-scale}
    \tau_{\mathrm{lib}}=\left(\frac{2\pi}{3|{C_{j}}|ne}\right)^{1/2}.
\end{equation}
Following \cite{OgiharaKobayashi2013}, we write the critical migration time-scale $\tau_{\mathrm{a,crit}}$ as:
\begin{equation}
    \label{eq:ta_crit}
    \tau_{\mathrm{a,crit}}=\frac{\tau_{\mathrm{lib}}a}{\Delta_a}=\left(\frac{2\pi}{16C_{j}^{2}e^2}\right)^{1/2},
\end{equation}
below which the corresponding MMR will break. In the case of convergent migration of two planets, the equilibrium eccentricities of planets balancing eccentricity-damping from the disc and excitation by resonant interaction reads:
\begin{equation}
    \label{eq:TP_ecc}
    e^2\sim \frac{\tau_{\mathrm{e}}}{\tau_{\mathrm{a}}}
\end{equation}
\citep{GoldreichSchlichting2014,TeyssandierTerquem2014,TerquemPapaloizou2019}. Combining \eq{ta_crit} and \eq{TP_ecc} we obtain:
\begin{equation}
    \label{eq:result}
    \tau_{\mathrm{a,crit}}\sim \frac{2\pi}{16C_{j}^2\tau_{\mathrm{e}}}.
\end{equation}
This expression implies that we can either employ efficient eccentricity-damping (low $\tau_{\mathrm{e}}$) or use smaller planet mass (low $C_{j}$), in order to enlarge $\tau_{\mathrm{a,crit}}$ and cross the MMR. Note that although the above expressions are derived for an internal MMRs (where the outer planet is fixed), those for the external MMRs are almost the same \citep{MurrayDermott1999}.

The above analytic results is consistent with \fg{model4_sensitivity}. The value of $\tau_{a,\earth}$ at the boundary between triangles and circles indicate the critical migration time-scale $\tau_{a,\earth}$ of the 2:1 MMR of planet b \& c. As $C_e$ decreases or $A_e$ increases, the boundary moves to the right, implying longer $\tau_{a,\earth}$.

\subsection{Eccentricity-period relation in 3BRs}
\label{sec:commensurability}
In a three-planet system connected by one 3BR with each of the adjacent two planets connected by $j{-}1{:}j$ and $k-1{:}k$ MMRs, the eccentricities and period ratios are related to each other.
We define the distance to exact commensurability of planet $i$ and $i+1$ as:
\begin{equation}
    \label{eq:delta}
    \delta_{i,i+1} = \frac{(j-1)P_{i+1}}{j P_i}-1,
\end{equation}
where $P_{i}$ is the orbital period of planet $i$. 
The MMR angles in a planet pair near $j-1{:}j$ MMR are:
\begin{equation}
    \label{eq:2b_angle}
    \phi_{i,i+1;i+X}=(j-1)\lambda_{i}-j\lambda_{i+1}+\varpi_{i+X}.
\end{equation}
where $X=0$ or $X=1$.
The time derivatives $\dot{\phi}_{i,i+1;i+X}$ average to zero if $\phi_{i,i+1;i+X}$ is librating. Taking the time derivative of \eq{2b_angle} and combining it with  \eq{delta}, we have:
\begin{equation}
    \label{eq:delta_p}
    \delta_{i,i+1}=-\dot{\varpi}_{i+X}\frac{P_{i+1}}{2\pi j}.
\end{equation}
A similar relation holds for planet $i+1\,\&\,i+2$. Without external perturbation the three planets are locked in common apsidal precession,  $\dot{\varpi}_{i} =\dot{\varpi}_{i+1} =\dot{\varpi}_{i+2}$. Then, the ratio of the distance to exact commensurability of two planet pairs is:
\begin{equation}
    \label{eq:distance}
    \frac{\delta_{i+1,i+2}}{\delta_{i,i+1}}= \frac{j P_{i+2}}{k P_{i+1}}\approx\frac{j}{k-1}.
\end{equation}
In Laplace resonances where $j=k$, the right hand term of \eq{distance} is larger than unity. \cite{RamosEtal2017} and \cite{TeyssandierLibert2020} insert Lagrange's planetary equation, $\dot{\varpi}=C_j \cos\phi/e$, into \eq{delta_p} and get:
\begin{equation}
    \label{eq:delta_ecc}
    \delta_{i,i+1}=\frac{1}{j} \alpha_{j}^{-1/2} f_j \frac{m_{i+1}}{M_{\star}}\frac{1}{e_{i}}
\end{equation}  
Combining \eq{distance} and \eq{delta_ecc}, we arrive at:
\begin{equation}
    \label{eq:distance_ecc}
    \frac{e_i}{e_{i+1}}
    = \frac{m_{i+1}f_j}{m_{i+2}f_k}\left(\frac{\alpha_k}{\alpha_j}\right)^{1/2}\frac{P_{i+2}}{P_{i+1}}
    \approx \frac{m_{i+1}f_j}{m_{i+2}f_k}\left[\frac{k(j-1)}{j(k-1)}\right]^{1/3}\frac{k}{k-1}.
\end{equation}
In Laplace resonances, the right hand term of \eq{distance_ecc} is larger than unity when planets have nearly the same mass. Therefore, outer planets tend to further depart from exact commensurability (\eq{distance}) and have smaller eccentricities (\eq{distance_ecc}). The eccentricities and the period ratios relationships of model \modeli, in which planets stay in first-order MMR, is revealed by \fg{model2_separate}. The deviations from exact commensurability increase and the eccentricities decrease from planet b to planet h. In the final stage of \modeliv (Stage~III in \fg{model4_tide}, planet b \& c and c \& d are near MMRs of order higher than one. Because the analysis above only holds for first-order resonances, the relationship between the eccentricity and period ratios can be seen for outer planets but not for planet b/c/d.

\section{Discussion}
\label{sec:discussion}
\subsection{Model Assessment}
\label{sec:discuss_models}
In this work, we have aimed to connect the present-day dynamical properties of the TRAPPIST-1 planets to their formation. We have assumed that the planets are formed at locations further away from their present orbits, were subject to inward (Type~I) migration, which was halted at the disc inner edge. Subsequently, the planets formed a chain of first-order MMRs, predominantly of 3:2 commensurability, and a plethora of three-body resonances (Stage~I). In the next Stage~II, which is either separated in time from Stage~I (\se{3BR_test}) or occurs contemporaneously with Stage~I (\se{new_model}), the two inner-most planets fell into the cavity, allowing their period ratios to expand due to the (weak and distant) torque from the disc. Finally, stellar tidal interactions after the demise of the gas disc (Stage~III) provided the final act in moving the planets into a configuration consistent with their present-day orbital periods, period ratios, eccentricities, and three-body resonance angles. 

\changed{The first successful model matching the observed TRAPPIST-1 configuration has Stage I, II, and III occurring separately (\se{3BR_test}). However, the model has several drawbacks: (i) The probability of a successful case is low. (ii) It requires rather high additional eccentricity-damping on the planet at the disc inner edge ($A_e \gg 10$). Intriguingly, such enhanced eccentricity-damping at the disc edge has recently been observed (but not quite understood) in hydrodynamical simulations of planet-disc interaction \citep{AtaieeKley2021}. Magnetospheric effects may also contribute towards a pileup of material near the cavity radius \citep{D'AngeloSpruit2010}, which would be the most natural explanation for the enhanced damping. Therefore, this model demands a very high surface density at the edge of the disc or some unknown mechanisms that would give rise to the anomalous efficient eccentricity-damping around the truncation radius. (iii) The physical origin of the one-sided negative torque on the planet in the disc cavity is not motivated. The torque required is usually ${\sim}10$ times stronger than the Lindblad torque calculated in \se{oneside}. Hence, for this scenario to work, the surface density of the gas needs to be boosted by a factor ${\sim}10$ near the cavity region. Even then, the probability of an outcome consistent with the TRAPPIST-1 system remains low.} %times denser disc at the Lindblad resonance location would exert such a torque on the planets in the cavity. Note that if a disc with a much higher surface density around truncation radius is feasible, this model can reproduce the TRAPPIST-1 analog self-consistently, except for its low probability.

\changed{Finally, we have investigated the Early Cavity Infall model (\se{new_model}) which features planets b and c entering the cavity before the outer planets park in resonance near the inner disc, combining Stage~I and Stage~II. In this scenario, we adopt a value for the Lindblad torque acting on planet c consistent with the disc mass. The key drawback of this model is that it introduces another parameter -- the time interval between planet formation ($\tau_\mathrm{pl}$) -- which needs to be tuned such that it is at least one-tenth of the time-scale over which planet c is being pushed in.\footnote{That is, $\tau_\mathrm{1s}/10 \lesssim \tau_a + \tau_\mathrm{pl}$. The factor 10 reflects the distance over which planet c migrates, which is about 1/10 of its semi-major axis. Also, $\tau_a \ll \tau_\mathrm{pl}$.} Only then can planet c migrate sufficiently far inwards to cross the locations of the 3BRs before their appearance. In addition, success is more likely for low values of the eccentricity damping parameter $C_e$. Nevertheless, this scenario is attractive for the following reasons: (i) the planet formation interval time ($\tau_\mathrm{pl}$) of ${\sim}10^5\,\mathrm{yr}$ is in line with the pebble-driven growth scenario that we have previously explored \citep{OrmelEtal2017,SchoonenbergEtal2019}. (ii) Other parameters are also natural or in line with the constraints from Objective I. No anomalous eccentricity damping is present ($A_e=1$), the value of $A_a$ is in line with theory \citep{LiuEtal2017}, and the values of $\tau_{a,\earth}$ that are required are overall consistent with the Objective I results. Hence, we favour this scenario to explain the architecture of the the TRAPPIST-1 system.} 
%analog is much higher

%Here, we \coch{give a clear physical motivation}{adopt the val $m=1$ mode } for the one-side negative torque on the planets in the cavity \coch{--}{} Lindblad torque, whose migration time-scale is \coch{on the order of}{about}
%This corresponding time-scale is about two magnitudes longer than the Type~I migration time-scale. 
%the migration time-scale of planet c in the cavity is less than on the order of ten times larger than planet formation interval\ccc{unclear phrasing}.  of several barrier three-body resonances, which can trap planet c, are crossed before their \coch{existences}{appearance}

%both Objective~I and Objective~II (\fg{model4_sensitivity}, \fg{param} and \fg{model_new})
\changed{In the Early Cavity Infall model (\fg{model_new}), efficient eccentricity-damping is preferred to prevent the outer planets, especially planet d, from crossing the desired resonances due to the stochastic nature of resonance trapping at high eccentricity. For the TRAPPIST-1 system, such high eccentricity-over-semi-major axis damping (low $C_e$) is also used in other literature studies, e.g., \cite{TeyssandierEtal2021} and \cite{MacDonaldEtal2021}.
The most straightforward explanation is that Type~I disc migration, being a complex phenomenon, operated less efficient in TRAPPIST-1, i.e., that $\gamma_I$ is smaller than what linear theory prescribes. In addition, subdued Type-1 migration would also ensure $A_a$ will be above the required threshold. A low $C_e$ parameter is also required in the K2-24 planetary system to reproduce its observed TTV signals \citep{TeyssandierLibert2020}. }

\subsection{Resonances Establishing and Breaking}
Three-body resonances play a crucial role in shaping these scenarios. A multi-planet resonant system as TRAPPIST-1 is characterized by many adjacent and non-adjacent three-body resonances, with some of the $\Phi_i$ librating, while others do not. This dynamical imprint strongly constrains the formation history of the TRAPPIST-1 system. In 3BRs, resonance locking limits the ability for period expansion. In the literature, certain mechanisms that have been invoked to separate two-planet systems by divergent migration, e.g., stellar tidal damping \citep{LithwickWu2012,BatyginMorbidelli2013} or magnetospheric rebound \citep{LiuEtal2017}. When, however, the planets are instead connected to other planets through three-body resonances, all planets participate in the expansion and the effects on --say-- the inner two planets are potentially limited. Specifically, for TRAPPIST-1 the $\tbr{c,d,e}{2,5,3}$ resonance, corresponding to the 3:2 and 3:2 MMR of planets c/d/e, naturally forms during the migration phase (Stage I), but greatly hinders the further period expansion of the inner planet system towards their observed period ratios. As this angle is currently not librating, we conclude that it must have been broken early, which would be achieved when planets b and c fell into the cavity and were repulsed by the gas disc. This allows the subsequent expansion of the inner planet system b/c/d towards their observed orbital period ratios.

This (temporary) decoupling of the inner planet system from the other TRAPPIST-1 planets was already employed in the study by \citet{PapaloizouEtal2018}, where initially two dynamically separate subsystems (b/c/d and e/f/g/h), each locked in their respective three-body resonances, dynamically converge into the present configuration under the action of stellar tidal forces only. As in that scenario stellar tides would primarily operate on the inner system, it naturally explains its expansion. However, \citet{PapaloizouEtal2018} do not motivate the origin of the two subsystems, their initial separations and the initial period ratios within each system. Indeed, the initial setup of these simulations follows from the formation stage. We find that during the expansion, planets may well enter the wrong three-body resonance (especially $\tbr{c,d,e}{2,5,3}$, $\tbr{c,e,f}{3,15,12}$ and $\tbr{c,d,e}{5,14,9}$; see \fg{param_study_model}, \fg{param} and \fg{model_new}), which stellar tides alone is too weak to break. This supports our scenario where the inner-most planets were pushed into the disc cavity.

\changed{To achieve a first-order resonant chain, we need to assume that planets c and g start in a 3:2 and 4:3 resonance (\fg{model4_sensitivity}). \cite{ColemanEtal2019} and \cite{BurnEtal2021} also initialize planet embryos close to each other such to avoid trapping in wider MMR like 2:1 and 3:2. A natural explanation is that these planets formed rapidly one-after-another, which is conceivable in the pebble-driven formation model \citep{SchoonenbergEtal2019, LinEtal2021}. }

\changed{In this work, we simplify the disc model without complicating disc-planets interactions via introducing, e.g., radial gravity \citep{NagasawaEtal2003,PanEtal2020} and stochastic torque \citep{ReinPapaloizou2009, PaardekooperEtal2013} induced by the disc and raised on planets. How much difference those mechanisms can make to 3BR trapping and escaping need to be addressed in the future. }

\subsection{Application to other systems}

\changed{In our model, the planet in the disc cavity can be repulsed inward slowly by Lindblad resonance from the external disc. Three-body resonance with two planets in the external disc can then halt the inward migration in the disc cavity. This mechanims therefore enables a configuration with the inner planets more separated from each other than the outer planets. Although three-body resonances are not ubiquitous in the Kepler data \citep{GoldbergBatygin2021}, we expect that configurations with separated inner planets near higer-order resonances may be connected by three-body resonances.}
%if such configurations are discovered.}\ccc{I don't understand the last sentence}

\changed{We give three examples. HD 158259 is a six planets system \citep{HaraEtal2020}. The period ratios are $P_\mathrm{c}/P_\mathrm{b}=1.57$, $P_\mathrm{c}/P_\mathrm{b}=1.51$, $P_\mathrm{c}/P_\mathrm{b}=1.53$, $P_\mathrm{c}/P_\mathrm{b}=1.51$ and $P_\mathrm{c}/P_\mathrm{b}=1.44$, where planet b and c are more separated than other adjacent plant pairs. Interestingly, planet c, d and e in HD 158259 are near 3:2 mean-motion resonance, similar to planet d, e and f in TRAPPIST-1. Based on our analysis in \se{3BR_test}, we predict that the three-body resonance angle $\Phi_\mathrm{b,d,e}=3\lambda_\mathrm{b}-15\lambda_\mathrm{d}+12\lambda_e$ in HD 158259 system is librating. This angle is equivalent to $\Phi_2$ in TRAPPIST-1 (\se{3BR_test} and \se{new_model}). }

\changed{Other systems, while not sharing the same period ratios configuration as TRAPPIST-1, do feature wider-spaced inner planets. }
Recently, \cite{LeleuEtal2021} found that the five outermost planets of the six-planet TOI-178 system constitute a 2:4:6:9:12 chain of Laplace resonances. The period ratio of the second-to-innermost planet is $1.69$. \changed{Since this period ratio is close to 5:3, it may indicate that planet b is connected by the outer planets through three-body resonance.}
%connecting planet b and two more outer planets is librating.}\ccc{???}
In the Kepler-80 system, the period ratios of the six planets are 3.1, 1.511, 1.518, 1.35, and 1.538 from inner to outer respectively \citep{Xie2013,MacDonaldEtal2021}.
\changed{Although the period ratio of 3.1 cannot be explained by the Lindblad torque mechanism, it would still be worthwhile to investigate whether a 3BR trapping may occur at this location.}
%This system features both modestly-separated planets (just outside 3:2) for which our model is appropriate as well as a single more widely separated innermost pair, but we need \uwave{additionally evaluate}\ccc{???} the three-body resonances between the period ratio from 1.7 to 3.1. 
\changed{To address what role 3BRs in these systems have played in shaping the configuration of these systems, the first step is to evaluate their strength by numerical simulations such as done in this work.}
%The increasingly growing exoplanet census offers us the opportunity to assess whether the planet system template proposed in this work is more widely applicable. And the key step is to evaluate the strength of 3BRs in different configurations.

\section{Conclusions}
\label{sec:conclusion}
We conducted N-body simulations to reconstruct the origin of the dynamical structure of planets in the TRAPPIST-1 system. Our model covers the stages directly after planet formation (${\sim}10^5\,\mathrm{yr}$) in the proto-planet disc, until the present age. Two objectives were pursued: the formation of a chain of first-order MMR (Objective I) and the evolution of the inner-most planets towards their present dynamical configuration (Objective II). \changed{In the Late Infall Model planets b and c entered the disc cavity at some time after all planets entering the resonant chain, and were repulsed by the one-sided Lindblad torque exerted by the exterior disc. Together with stellar tidal damping, this provided for the expansion of the inner sub-system towards its present configuration.
In the preferred Early Infall Model, planet b and c fall in the cavity before the outer planets entering the resonant chain. Then, planet c can migrate across the resonant locations in the cavity before their existence. }
%We then proposed one complete model that is capable to reproduce the observed configuration, according to the numerical study on trapping and escape conditions of these 3BRs.

We list our main findings from this study:
\begin{enumerate}
\item The 4:3 MMR of planet f \& g stands out as the closest resonance. In simple disc models planets f \& g fail to cross the 3:2 resonance without preventing other planets from obtaining a more compact configuration. The 4:3 MMR could be explained as an imprint of rapid sequential planet formation, whereafter f and g migrated convergently.

%formation location of planet g happened to lie within the 3:2 resonance location at the time when planet f entered the resonance chain
%-- a scenario marginally . .

\item Simple scenarios that feature only disc and stellar tidal damping fail to meet Objective II. Due to three-body resonance locking, the inner subsystem cannot expand independently from the rest of the system.  
%We examined three relatively simple models and preliminarily concluded that the b/c/d subsystem needs to be separated enough to avoid the impediment from some 3BRs, on the way to the observed configuration. 
\item \changed{Objective II can be reached when planets b and c entered the disc cavity. The disc further repulsed the planets in the cavity by virtue of the order $m=1$ outer Lindblad torque whose resonant location still resides in the disc. This repulsion is a crucial mechanism in aligning the inner system towards the observed 3BRs and period ratios.}
\item \changed{Efficient eccentricity-damping ($C_e\sim0.1$)} \changedd{and a strong disc inner barrier ($A_a\sim100$) were found to be advantageous to obtain the desired resonance configuration.} \changedd{These imply a small value for the Type~I migration prefactor $\gamma_I$.}
%compared to \cite{TanakaWard2004} and \cite{CresswellNelson2008}), for both satisfying Objective~I and Objective~II. It mainly refrain planet d and e cross 3:2 MMR during the disc phase. \ccc{unclear to me}}
\item \changed{In the Late Infall model, planet b and c enter the disc cavity after outer planets enter the resonant chain. To reproduce TRAPPIST-1, it is necessary to enhance the eccentricity damping of planet d at the cavity by large amounts ($A_e\gtrsim10$).}
%on the planets at the disc inner edge much more efficient. }
\item \changed{In our preferred scenario, planet b and planet c enter the disc cavity early, before the outer planets have joined the resonance chain. To allow planet c to migrate across the barrier resonances, the planet formation interval time $\tau_\mathrm{pl}$ is about $10^5\mathrm{yr}$.}%, which is consistent with \cite{OrmelEtal2017} and \cite{SchoonenbergEtal2019}. }
\item Tidal dissipation decreases the eccentricities and increases the period ratio among planets in a three-body resonance chain.
After all planets have been connected by three-body resonances, it is the period ratio of the outermost planet pair (h and g) that increases most, while the rate of the expansion is determined by the tidal dissipation of the innermost planet b.
For our model, we constrain the tidal quality factor of planet b $Q_b\gtrsim200k_2$.
\item Three-body resonances play an important role in shaping the dynamical structure of the TRAPPIST-1 planets. In contrast to two-body resonances, 3BRs do not necessarily break when exposed to a divergent force. In this study, a numerical approach was used to assess the strength of 3BR and we have highlighted the role of non-adjacent 3BR. 
%We also find that more efficient eccentricity-damping promotes resonance crossing when eccentricities is low ($\lesssim0.01$) \ccc{maybe confusing}, and provides an analytical motivation (\se{MMR_ecc}). 
Due to their significance in shaping scenarios containing divergent migration, a more comprehensive understanding of 3BR is in need.
\end{enumerate}

\section*{Acknowledgements}
We thank the referee for their constructive suggestions and comments. We thank Eric Agol for providing us with an early version of the transit timing and photometric analysis of TRAPPIST-1 planets \citep{AgolEtal2021} and corrections. We also thank Ramon Brasser, Beibei Liu and Jean Teyssandier for useful email discussions. 
Software: \texttt{Matplotlib} \citep{Hunter2007,CaswellEtal2021}, \texttt{REBOUND} \citep{ReinLiu2012} and \texttt{REBOUNDx} \citep{TamayoEtal2020}. We also use \texttt{Pylaplace} to calculate the Laplace coefficients, \url{https://pypi.org/project/pylaplace/}.

%%%%%%%%%%%%%%%%%%%%%%%%%%%%%%%%%%%%%%%%%%%%%%%%%%
\section*{Data Availability}
The data underlying this article will be shared on reasonable requests
to the corresponding author.
%The inclusion of a Data Availability Statement is a requirement for articles published in MNRAS. Data Availability Statements provide a standardised format for readers to understand the availability of data underlying the research results described in the article. The statement may refer to original data generated in the course of the study or to third-party data analysed in the article. The statement should describe and provide means of access, where possible, by linking to the data or providing the required accession numbers for the relevant databases or DOIs.

%%%%%%%%%%%%%%%%%%%% REFERENCES %%%%%%%%%%%%%%%%%%

% The best way to enter references is to use BibTeX:

\bibliographystyle{mnras}
\bibliography{ads} % if your bibtex file is called example.bib

% Alternatively you could enter them by hand, like this:
% This method is tedious and prone to error if you have lots of references
%\begin{thebibliography}{99}
%\bibitem[\protect\citeauthoryear{Author}{2012}]{Author2012}
%Author A.~N., 2013, Journal of Improbable Astronomy, 1, 1
%\bibitem[\protect\citeauthoryear{Others}{2013}]{Others2013}
%Others S., 2012, Journal of Interesting Stuff, 17, 198
%\end{thebibliography}

%%%%%%%%%%%%%%%%%%%%%%%%%%%%%%%%%%%%%%%%%%%%%%%%%%

%%%%%%%%%%%%%%%%% APPENDICES %%%%%%%%%%%%%%%%%%%%%

\newpage
\appendix

\section{Lindblad torque}
\label{sec:Ward_calcu}
Following \citet{Ward1997}, we calculate the magnitude of the Lindblad torque at different order ($m$) of Lindblad resonance. We insert disc scale height $h=0.03$, disc power-law surface density gradient $k=d\mathrm{ln}\Sigma_\mathrm{g}/d\mathrm{ln}r=0.5$ and disc temperature gradient $l=\mathrm{d}\ln T/\mathrm{d}\ln r=1$. The left panel of \fg{lindblad} shows the Lindblad torque scale by Type~I migration torque at the first 50 orders Lindblad resonance location and the right panel only shows the first five orders. For m=1 only an outer Lindblad torque is present, which is used in this work. Its value is 0.0054.

\begin{figure}
    \centering
    \includegraphics[width=1\columnwidth]{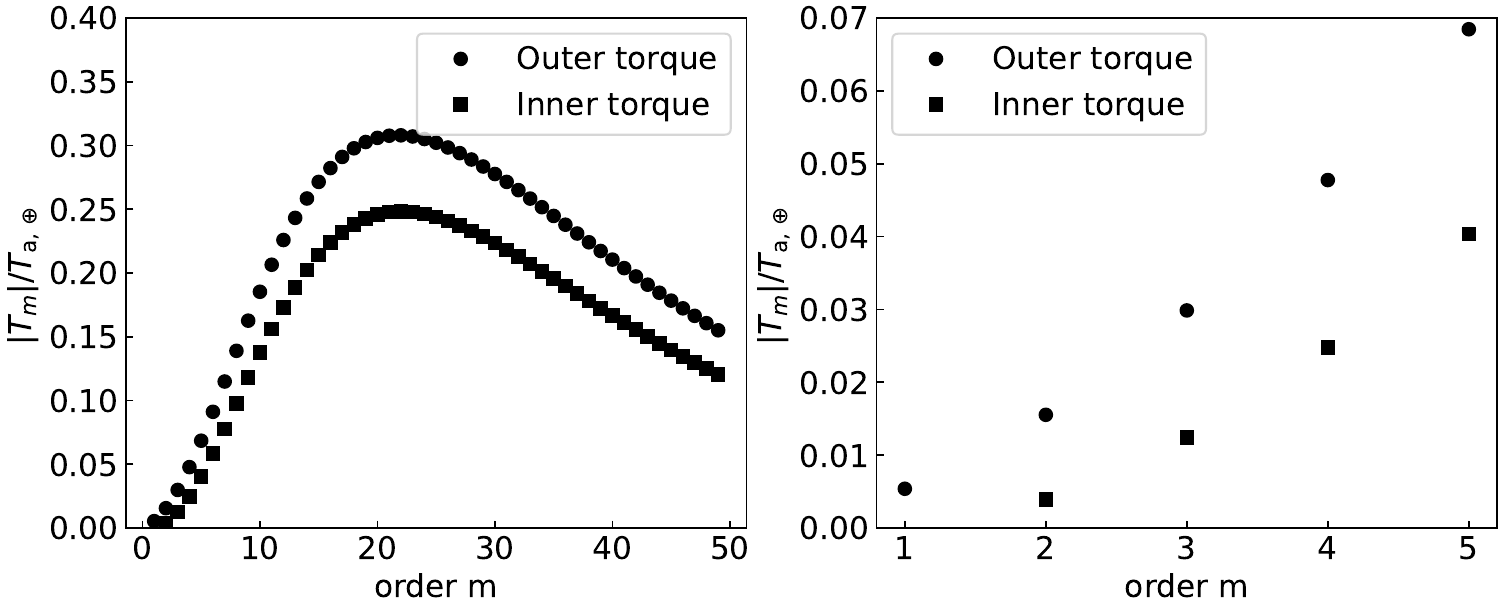}
    \caption{Torque magnitude as a function of order, m, normalized to the Type~I migration time-scale for scale height $h=0.03$ and disc surface density gradient $k=0.5$. Left panel present the order m from 1 to 50. Right panel zoom in m from 1 to 5. }
    \label{fig:lindblad}
\end{figure}

% \begin{table}
% 	\centering
% 	\caption{Torque magnitude as a function of order, m, normalized to the Type~I migration time-scale for scale hight $h=0.03$ and disc surface density gradient $k=0.5$.}
% 	\label{tab:Ward_calcu}
% 	\begin{tabular}{lccccccc} % four columns, alignment for each
% 		\hline
%          m=  & Outer   & Inner \\
%         \hline
%          1   & 0.0054  & - \\
%          2   & 0.0155  & -0.0039\\
%          3   & 0.0299  & -0.0124\\
%          4   & 0.0478 & -0.0248\\
% 		\hline
% 	\end{tabular}
% \end{table}

%%%%%%%%%%%%%%%%%%%%%%%%%%%%%%%%%%%%%%%%%%%%%%%%%%

% Don't change these lines
\bsp	% typesetting comment
\label{lastpage}
\end{document}